\documentclass[12pt,showpacs,preprintnumbers,superscriptaddress,amsmath,amssymb,nofootinbib]{revtex4}
\usepackage{graphicx}
\usepackage{dcolumn}
\usepackage{bm}
\usepackage{amssymb}
\usepackage{amsmath}
\usepackage{epsfig}    
\usepackage{color}
\usepackage{slashed}
\usepackage{float}
\begin{document} 

\title{Perturbative Unitarity Bounds \\ in  Composite 2-Higgs Doublet Models}

%
\author{Stefania De Curtis}
\email{decurtis@fi.infn.it}
\affiliation{
INFN, Sezione di Firenze, and Department of Physics and Astronomy, University of Florence, Via
G. Sansone 1, 50019 Sesto Fiorentino, Italy}
\author{Stefano Moretti}
\email{S.Moretti@soton.ac.uk}
\affiliation{School of Physics and Astronomy, University of Southampton, Southampton, SO17 1BJ, United Kingdom}
\author{Kei Yagyu}
\email{K.Yagyu@soton.ac.uk}
\affiliation{School of Physics and Astronomy, University of Southampton, Southampton, SO17 1BJ, United Kingdom}
\author{Emine Yildirim}
\email{ey1g13@soton.ac.uk}
\affiliation{School of Physics and Astronomy, University of Southampton, Southampton, SO17 1BJ, United Kingdom}

\begin{abstract}
\noindent
We study bounds from perturbative unitarity in a Composite 2-Higgs Doublet Model (C2HDM)
based on the spontaneous breakdown of a global symmetry $SO(6)\to SO(4)\times SO(2)$ at the compositeness scale $f$. 
The eight pseudo Nambu-Goldstone Bosons (pNGBs) emerging from such a dynamics are identified as  two isospin doublet Higgs fields.  
We calculate the $S$-wave amplitude for all  possible 2-to-2-body elastic (pseudo)scalar boson scatterings at energy scales $\sqrt{s}$ reachable at the Large Hadron Collider (LHC) and beyond it,   
including the longitudinal components of weak gauge boson states as the corresponding pNGB states.  
In our calculation, the Higgs potential is assumed to have  the same form as that in the Elementary 2-Higgs Doublet Model (E2HDM)  
with a discrete $Z_2$ symmetry, which is expected to be generated at the one-loop level via the Coleman-Weinberg (CW) mechanism. 
We find that the $S$-wave amplitude matrix can be block-diagonalized with maximally $2\times 2$ submatrices
in a way similar to  the E2HDM case
as long as we only keep the contributions from ${\cal O }(\xi s)$ and ${\cal O }(\xi^0 s^0)$ in the amplitudes, where $\xi=v_{\text{SM}}^2/f^2$ and $v_{\text{SM}}^{}\simeq 246$ GeV, which is an appropriate approximation for our 
analysis. 
By requiring the C2HDM to satisfy perturbative unitarity at energies reachable by the LHC,  we derive bounds on its parameters such as
$\xi$ and the masses of extra Higgs bosons present in the scenario alongside the Standard Model (SM)-like Higgs state discovered in 2012. 

\end{abstract}
\maketitle

\section{Introduction}

The search for additional Higgses, after the one discovered so far \cite{Aad:2012tfa,Chatrchyan:2012ufa} and the possible evidence of a  new (pseudo)scalar state with mass around 750~GeV \cite{Diphoton}, is one of the most important tasks of Run 2 of the Large Hadron Collider (LHC). It is widely known that extra spinless states  with or without Standard Model (SM) quantum numbers can induce sizeable tree-level effects in the couplings
of the discovered state, which have been under close scrutiny for three years now. It is  also true that direct searches
for new Higgs states, as shown by the aforementioned recent preliminary results, could have a dramatic impact on
LHC activities. These two facts  seem already  remarkable motivations to study the phenomenology of extra Higgses at the present CERN machine.

Despite the obvious far-reaching consequences of a discovery of even a single additional scalar, the presence of another Higgs would not be, by itself, an evidence for the naturalness of the weak scale: such a defining situation would still be pending upon the whole subject. Just like for the case of a single Higgs doublet,  for which the hierarchy problem can be explained by its pseudo Nambu-Goldstone Boson (pNGB) nature, we would like to link the presence of extra Higgs particles to natural theories of the Fermi scale.  In particular, we have in mind composite Higgs models, where the mass of the lightest Higgs state is kept naturally lighter than a new strong scale around $\sim $ TeV by an approximate global symmetry \cite{Dugan:1984hq}, broken by SM interactions in the \textit{partial compositeness} paradigm \cite{Kaplan:1991dc,Contino:2006nn}. 

In the minimal composite Higgs model \cite{MCHM, Contino:2006qr}, the only light scalar in the spectrum is a pNGB, surrounded by spanned composite resonances roughly heavier by a loop factor. The underlying symmetries protect the Higgs mass from quantum corrections thus giving a simple solution to the hierarchy problem. 
The only robust way to expect new light (pseudo)scalars in the spectrum  is to make them also pNGBs. Even in the case they are not expected to be as light as the SM Higgs, it is interesting to find a mechanism for describing all the Higgses as pNGBs and to explain their mass differences.  Last, but not least for importance, in the case of  extra Higgs doublets with no Vacuum Expectation Value
(VEV) nor couplings to quark and leptons, one could also have the possibility to describe neutral light states as possible  composite dark matter candidates \cite{IDM}.

In this paper we aim at  identifying among the lightest scalars at least two Higgs doublets as this would lead to a Composite 2-Higgs Doublet Model (C2HDM) \cite{so6}. 
The latter represents the simplest natural two Higgs doublet alternative to supersymmetry. The composite Higgses arising from a new dynamics at the TeV scale ultimately drive the Electro-Weak (EW) symmetry breaking.
To include them as pNGBs, one has basically two different and complementary approaches: (i)
to write down an effective Lagrangian (e.g., \textit{a la} Strongly Interacting Light Higgs (SILH) \cite{SILH}) invariant under SM
symmetries  for  light composite $SU(2)$ Higgses; (ii) to explicitly impose a specific symmetry breaking structure containing multiple pNGBs. We take here the second approach. In particular, we will study in detail models based on  the spontaneous global symmetry breaking of $SO(6)\to SO(4)\times SO(2)$~\cite{so6}. We will focus on their predictions for the structure of the (pseudo)scalar spectrum and the deviations of their couplings from those of a generic renormalizable Elementary
2-Higgs Doublet Model
 (E2HDM). In the $f\to \infty$ limit the  pNGB states are in fact identified with the physical Higgs states of doublet scalar fields of the E2HDM\footnote{For an updated review of the theory and phenomenology of E2HDMs see~\cite{branco}. }. Deviations from the E2HDM are parametrized by $\xi=v_{\rm SM}^2/f^2$, with $v_{\rm SM}$ the SM Higgs VEV.

Once the strong  sector is integrated out, the pNGB Higgses, independently of their microscopic origin, are  described by a non-linear $\sigma$-model associated to the coset.
We construct their effective low-energy Lagrangian  according to the prescription developed by Callan, Coleman, Wess and Zumino (CCWZ) \cite{ccwz}, which makes only few specific assumptions about the strong sector, namely, the global symmetries, their pattern of spontaneous breaking and the sources of explicit breaking (we assume that they come from the couplings of the strong sector with the SM fields).
The scalar potential is in the end generated by loop effects and, at the lowest order, is mainly determined by the free parameters associated to the top sector \cite{so6}. 

Here we will focus on the  unitarity properties of a C2HDM\footnote{{For the discussion of unitarity in minimal composite Higgs models see \cite{Kanemura:2014kga}.} }, namely, we will derive the bounds on the parameters of the model by requiring perturbative unitarity to hold at the energies reachable by the LHC. In fact, contrarily to the E2HDM, which is renormalizable, the C2HDM is an effective theory.  The pNGB nature of the Higgses leads to a modification of their couplings to matter with respect to the E2HDM case and, as a consequence, forces to a non-vanishing $s$-dependence of the scattering amplitudes. This means that the C2HDM is not unitary for energies above a critical value or, alternatively said, one needs to consider other new physics contributions (e.g., new composite fermions and gauge bosons)
to make the model unitary above that energy scale.   Since the fermion content of the model is not playing a role in the present investigation, we will not specify the fermion representation and we will not calculate the Higgs potential generated by the radiative corrections. Instead, we will assume the same general form of the Higgs potential as in the E2HDM with a $Z_2$ symmetry,  {the latter imposed in order to avoid Flavor Changing Neutral 
Currents (FCNCs) at the tree level~\cite{GW}.} Therefore, in
 the energy region where the E2HDM and  C2HDM are both unitary, it is interesting to compare the bounds on the additional Higgs masses. In fact, due to a compensation amongst mass- and energy-dependent contributions, we find that regions not allowed in the E2HDM are instead permitted in the C2HDM for the most general configuration of their parameter spaces. 

The paper is organized as follows. In Section~II we describe the C2HDM based on $SO(6)/SO(4)\times SO(2)$.  We separately discuss two scenarios:  the active one in which both Higgs doublet fields acquire a VEV and the inert one in which only one does.
In Section~III, the unitarity properties of the C2HDM are discussed by calculating all the 2-to-2-body (pseudo)scalar boson scattering amplitudes and by deriving constraints through all these channels.
Conclusions are drawn in Section~IV. Some  technical details of the derivation of the pNGB kinetic terms are given in Appendix A.

\section{The Model}

\subsection{Higgs Doublets as pNGBs}

We first discuss how we obtain two isospin scalar doublets from the spontaneous breakdown of the global symmetry, i.e., $SO(6)\to SO(4)\times SO(2)$. 
In order to clarify this, we introduce the following fifteen $SO(6)$ generators:
\begin{align}
&T_{L,R}^a = -\frac{i}{2}\left[\frac{1}{2}\epsilon^{abc}(\delta_i^b\delta_j^c-\delta_j^b\delta_i^c)\mp (\delta_i^a\delta_j^4-\delta_j^a\delta_i^4)\right], 
\notag\\
&T_S= -\frac{i}{\sqrt{2}}(\delta_i^5\delta_j^6-\delta_j^5\delta_i^6), \notag\\ 
&T_1^{\hat{a}} = -\frac{i}{\sqrt{2}}(\delta^{\hat{a}}_i \delta^5_j - \delta^{\hat{a}}_j \delta^5_i ), \quad 
T_2^{\hat{a}} = -\frac{i}{\sqrt{2}}(\delta^{\hat{a}}_i \delta^6_j - \delta^{\hat{a}}_j \delta^6_i ),\notag\\
&\text{with}~~(a,b,c=1\text{-}3),~~(i,j=1\text{-}6),~~(\hat{a}=1\text{-}4).   \label{so6_gen}
\end{align}
The above generators are classified into the seven unbroken generators $T_{L,R}^a$ and $T_S$ 
and the eight broken generators $T_1^{\hat{a}}$ and $T_2^{\hat{a}}$. 
We can confirm that $T_{L,R}^a$ and $T_S$ are the subalgebras which generate the $SO(4)\times SO(2)$ subgroup by looking at the following commutation relations:
\begin{align}
&[T_L^a,T_L^b] =i\epsilon^{abc}T_L^c,\quad 
[T_R^a,T_R^b] =i\epsilon^{abc}T_R^c,\quad 
[T_L^a,T_R^b] = [T_L^a,T_S] = [T_R^a,T_S]=0,  \label{xx} \\
&[T_L^a,T_{\Phi_\alpha}] = -\frac{1}{2}\sigma^a T_{\Phi_\alpha},\quad 
[T_R^3,T_{\Phi_\alpha}] = -\frac{1}{2} T_{\Phi_\alpha}, \label{yy}
\end{align}
where 
\begin{align}
T_{\Phi_\alpha} = \begin{pmatrix}
T_\alpha^2 +i T_\alpha^1 \\
T_\alpha^4 -i T_\alpha^3 
\end{pmatrix},\quad \alpha = 1,2. 
\end{align}
Eq.~(\ref{xx}) tells us that the commutation relations among $T_{L,R}^a$ and $T_S$ are closed plus that
$T_L^a$ ($T_R^a$) generates the $SU(2)_L$ ($SU(2)_R$) subgroup of $SO(4)$ which is  identified as the custodial symmetry of the SM Higgs sector. 
Furthermore, Eq.~(\ref{yy}) shows that $T_{\Phi_\alpha}$ transforms as the $SU(2)_L$ doublet with  charge $+1/2$\footnote{The overall minus sign is a conventional as the $T_R^3$ charge should be +1/2 to get $Y=+1/2$ and $Q=+1$ for the upper component of the Higgs doublet.}. Therefore  the broken generators are associated with the pNGBs, transforming as a ({\bf 4},{\bf 2}) of $SO(4)\times SO(2)$.

We then introduce  the following two $SU(2)_L$ doublet scalar fields
associated with $T_{\Phi_\alpha}$ as pNGBs:
\begin{align}
\Phi_\alpha \equiv 
\frac{1}{\sqrt{2}}
\begin{pmatrix}
h_\alpha^2 +i h_\alpha^1 \\
h_\alpha^4 -i h_\alpha^3 
\end{pmatrix}
\equiv
\begin{pmatrix}
\omega_\alpha^+ \\
\frac{v_\alpha + h_\alpha + i z_\alpha}{\sqrt{2}}
\end{pmatrix},   \label{weakeigen}
\end{align}
where the $v_\alpha$'s are the VEVs of $\Phi_\alpha$. 
The relation among $v_1$, $v_2$ and the Fermi constant $G_F$ will be discussed in Section~II-B. 
Notice that, in order to assign the right hypercharge to fermions, one has to introduce also an extra $U(1)_X$. The electric charge $Q$ will be then defined as usual by $Q=T_L^3+Y$ with 
the hypercharge $Y$  given by $Y=T_R^3+X$ where $X$ is the $U(1)_X$ charge. 
In this paper, we do not discuss the fermion sector which is not relevant to the following analysis, so we can omit to include this extra $U(1)_X$ group.

\subsection{Kinetic Lagrangian}

In general, once the coset space has been chosen, the low-energy Lagrangian is fixed at the two-derivative level,  the basic ingredient being the pNGB matrix which transforms non-linearly under the global group.  

The kinetic Lagrangian invariant under the $SO(6)$ symmetry 
can be constructed by the analogue of the construction in non-linear sigma models
developed in \cite{ccwz}, as 
\begin{align}
\mathcal{L}_{\text{kin}}=\frac{f^2}{4}(d_\alpha^{\hat{a}})_\mu(d_\alpha^{\hat{a}})^\mu, \label{ccwz1}
\end{align}
where 
\begin{align}
(d_\alpha^{\hat{a}})_\mu = i\,\text{tr}(U^\dagger D_\mu U T_\alpha^{\hat{a}} ).  \label{ccwz2}
\end{align}
Here $U$ is the pNGB matrix:
\begin{align}
&U=\exp\left(i\frac{\Pi}{f}\right),~~\text{with}~~
\Pi\equiv\sqrt{2}h_{\alpha}^{\hat{a}}T^{\hat{a}}_{\alpha}=-i\left( \begin{array}{ccc} 0_{4\times 4} & h_1^{\hat{a}} & h_2^{\hat{a}}\\
-h_1^{\hat{a}} & 0  & 0 \\
-h_2^{\hat{a}} & 0 & 0 \end{array} \right). 
\label{u6}
\end{align}
In Eq.~(\ref{ccwz2}), the covariant derivative $D_\mu$ is given by 
\begin{align}
D_\mu &= \partial_\mu -ig T_L^a W_\mu^a -ig' YB_\mu. 
\end{align}
The expressions for $(d_\alpha^{\hat{a}})_\mu$ up to ${\cal O}(1/f)$ are given in Appendix~A. 

In order to see how the gauge boson masses are generated, let us consider the fourth components of the Higgs fields:
\begin{align}
h_1^{\hat{a}} =  (0,0,0,\tilde{h}_1),\quad
h_2^{\hat{a}} = (0,0,0,\tilde{h}_2), \label{lim}
\end{align}
with 
\begin{align}
\tilde{h}_1 = h_1 + v_1 ,\quad \tilde{h}_2 = h_2 + v_2. 
\end{align}
In this case, the matrix $U$ defined in Eq.~(\ref{u6}) takes a simple form,
\begin{align}
U &=  \begin{pmatrix}
1_{4\times 4} -(1-\cos \frac{\tilde{h}}{f})_{44}  &  \frac{\tilde{h}_1^{\hat{a}}}{\tilde{h}}\sin\frac{\tilde{h}}{f} & \frac{\tilde{h}_2^{\hat{a}}}{\tilde{h}}\sin\frac{\tilde{h}}{f} \\
-\frac{\tilde{h}_1^{\hat{a}}}{\tilde{h}}\sin\frac{\tilde{h}}{f}  & 1-\frac{\tilde{h}_1^2}{\tilde{h}^2}(1-\cos \frac{\tilde{h}}{f}) & -\frac{\tilde{h}_1 \tilde{h}_2}{\tilde{h}^2}(1-\cos \frac{\tilde{h}}{f}) \\
-\frac{\tilde{h}_2^{\hat{a}}}{\tilde{h}}\sin\frac{\tilde{h}}{f} & -\frac{\tilde{h}_1 \tilde{h}_2}{\tilde{h}^2}(1-\cos \frac{\tilde{h}}{f}) &1- \frac{\tilde{h}_2^2}{\tilde{h}^2}(1-\cos \frac{\tilde{h}}{f})
\end{pmatrix},  \label{u62}
\end{align} 
where 
\begin{align}
\tilde{h}\equiv \sqrt{\tilde{h}_1^2+\tilde{h}_2^2}.  
\end{align}
The 2-gauge boson terms are extracted from Eqs.~(\ref{ccwz1}), (\ref{ccwz2}) and (\ref{u62}) as  
\begin{align}
{\cal L}^{\text{mass}}_{\text{kin}} &= \frac{f^2}{8}(2g^2W_\mu^+ W^{-\mu} + g_Z^2Z_\mu Z^\mu)\sin^2\frac{\tilde{h}}{f}, 
\end{align}
and thus the gauge boson masses are given by  
\begin{align}
m_W^2 &= \frac{g^2}{4}f^2  \sin^2\frac{v}{f}, \quad
m_Z^2 = \frac{g_Z^2}{4}f^2\sin^2\frac{v}{f},
\end{align}
where $v^2\equiv v_1^2 + v_2^2$ and $g_Z^{}=g/\cos\theta_W$ with $\theta_W$ being the weak mixing angle.   
Notice here that the VEV $v$ is different from the one $v_{\text{SM}}^{}$ in the SM as long as we take a finite value of $f$. 
The relationship among $v$, $v_{\text{SM}}$ and $G_F$ is expressed as follows
\begin{align}
v_{\text{SM}}^2\equiv \frac{1}{\sqrt{2}G_F } = f^2\sin^2\frac{v}{f} \simeq (246~\text{GeV})^2.  \label{vsm}
\end{align}
The ratio of  the two VEVs is defined by $\tan\beta = v_2/v_1$. 

Similarly to the  E2HDM, we can define the so-called Higgs basis~\cite{HB} in which only one of the two doublet fields contains the VEV $v$ and the
{Nambu-Goldstone} states 
$G^\pm$ and $G^0$  absorbed into the longitudinal components of $W^\pm$ and $Z$ bosons, respectively:
\begin{align}
\begin{pmatrix}
\Phi_1 \\
\Phi_2
\end{pmatrix} 
=R(\beta)
\begin{pmatrix}
\Phi \\
\Psi
\end{pmatrix} ,\quad  R(x)= \begin{pmatrix}
\cos x & -\sin x \\
\sin x & \cos  x 
\end{pmatrix},
\end{align}
where 
\begin{align}
\Phi = \begin{pmatrix}
G^+ \\
\frac{v+h_1' + iG^0}{\sqrt{2}}
\end{pmatrix}, \quad 
\Psi = \begin{pmatrix}
H^+ \\
\frac{h_2' + iA}{\sqrt{2}} 
\end{pmatrix}. 
\end{align}
In the above expressions, $H^\pm$ and $A$ 
are the physical charged and CP-odd neutral Higgs boson, respectively, while 
$h_1'$ and $h_2'$ are the CP-even Higgs bosons which in general can be mixed with each other. 
In this basis, the two-derivative terms for scalar bosons are extracted up to ${\cal O}(1/f^2)$: 
\begin{align} 
\hspace{-5mm}{\cal L }_{\text{kin}}^{\text{2-der}} &= \left(1-\frac{\xi}{3}\right)\left[
|\partial_\mu G^+|^2 + \frac{1}{2}(\partial_\mu G^0)^2 + \frac{1}{2}(\partial_\mu h_2')^2 \right]
 +|\partial_\mu H^+|^2 + \frac{1}{2}(\partial_\mu A)^2 + \frac{1}{2}(\partial_\mu h_1')^2, 
\end{align}
where 
\begin{align}
\xi = \frac{v_{\text{SM}}^2}{f^2}. 
\end{align}
We see that the kinetic terms for $G^\pm$, $G^0$ and $h_2'$ 
are not in canonical form and we need to rescale the fields:
\begin{align}
G^+ \to \left(1-\frac{\xi}{3}\right)^{-1/2}G^+,\quad
G^0 \to \left(1-\frac{\xi}{3}\right)^{-1/2}G^0,\quad
h_2' \to \left(1-\frac{\xi}{3}\right)^{-1/2}h_2'. \label{shifts}
\end{align}
After this shift, we can define 
the mass eigenstates for the CP-even scalar bosons by introducing the mixing angle $\theta$ as:  
\begin{align}
\begin{pmatrix}
h_1' \\
h_2'
\end{pmatrix}
=
R(\theta)
\begin{pmatrix}
h \\
H
\end{pmatrix}, \label{theta} 
\end{align}
where $h$ is assumed to be the observed Higgs boson with a  mass of 125 GeV. 
{The mixing angle $\theta$ is determined by the mass matrix for the CP-even states calculated from 
the Higgs potential, which will be discussed in the next subsection.}

\subsection{Higgs Potential}

The Higgs potential is generated through the Coleman-Weinberg (CW) mechanism~\cite{cw} at loop levels. 
There are two types of  contributions to the potential, coming from gauge boson loops and fermion loops. 
The former contribution can be calculated without any ambiguities and it  generates a positive squared mass term in the potential~\cite{MCHM}. 
Thus, EW symmetry breaking does not occur by the gauge loops alone. Fermion loops can provide a negative contribution to the squared mass term, so their effect is essentially important to trigger EW symmetry breaking. However, the contribution from fermion loops depends on the choice of the representation of fermions.

The structure of the Higgs potential  in the $SO(6)/SO(4)\times SO(2)$ model has been studied in Ref.~\cite{so6} assuming several representations of fermion fields. They also assume that the explicit breaking of the global symmetry is associated with the couplings of the strong sector to the SM fields, that is, gauge and Yukawa interactions. This assumption, dictated by minimality, allows  one to parameterize the Higgs potential, at each given order in the fermion and gauge couplings, in terms of a limited number of coefficients. If this assumption is relaxed, the parameter space of the C2HDM could be significantly enlarged. The form of the potential they obtain is given by the general E2HDM one, but 
each of the parameters is expressed in terms of those in the strong sector (mainly associated to the top dynamics).
In our paper, however, we do not explicitly calculate the CW potential and we do not  specify the fermion representations, making our analysis  applicable  to different choices of them. In fact, while the coupling of the vector bosons is fixed by  gauge invariance, more freedom exists in the fermion sector and, to specify the model, one must fix the quantum numbers of the strong sector operators which mix with the SM fermions, in particular with the top quark. The CW potential  clearly depends on these choices.
Instead of performing the explicit calculation, we assume here the same form of the Higgs potential as that in the  E2HDM. 
Our results on the unitarity properties of the C2HDM   will be expressed as bounds on the masses of the Higgses  which are free parameters in the E2HDM. Once explicitly specified the model, we will have the possibility to check, by calculating the CW potential, if the composite Higgs spectrum of that particular configuration satisfies the unitarity bounds.

In order to avoid FCNCs at the tree level, a discrete $Z_2$ symmetry~\cite{GW} is often imposed onto the potential, which is what we also do here\footnote{In Ref.~\cite{so6}, 
the $Z_2$ symmetry ($\Phi_1\to +\Phi_1$ and $\Phi_2 \to -\Phi_2$) is referred as the $C_2$ symmetry 
whose transformation can be expressed by a diagonal $6\times 6$ matrix form acting on the $6\times 6$ pNGB matrix given in Eq.~(\ref{u6}).}. 
Under the $Z_2$ symmetry, 
the two doublet fields are transformed as $(\Phi_1,\Phi_2)\to (+\Phi_1,-\Phi_2)$. 
This symmetry can also avoid a large contribution to the EW $T$-parameter which could emerge in C2HDMs from  
the dimension 6 operator in the kinetic Lagrangian\footnote{The issue of anomalous contribution to the $T$-parameter and to FCNCs in C2HDMs is faced also 
in \cite{enrico} where they discuss $T$-safe models based on different cosets, in particular SO(9)/SO(8). }. 
Depending on the nature of the $Z_2$ symmetry, i.e., softly-broken or unbroken, the properties of the Higgs bosons can drastically change. 
In the following, we first discuss the softly-broken $Z_2$ case and then we consider the unbroken case. 
For the latter case, the VEV of $\Phi_2$ must be taken to be zero to avoid the spontaneously breakdown of the $Z_2$ symmetry. In analogy with the E2HDM we will refer to the former scenario as the active C2HDM, while the latter describes the inert C2HDM.
 
\subsubsection{Active C2HDM}

The Higgs potential under the gauge symmetry $SU(2)_L\times U(1)_Y$ with the softly-broken $Z_2$ symmetry is given by 
\begin{align}
V(\Phi_1, \Phi_2)&=
m_1^2 \Phi_1^{\dagger}\Phi_1 + m_2^2 \Phi_2^{\dagger}\Phi_2 
- m_3^2 (\Phi_1^{\dagger}\Phi_2 + \text{h.c.})
+ \frac{1}{2}\lambda_1 (\Phi_1^{\dagger} \Phi_1)^2 + \frac{1}{2}\lambda_2 (\Phi_2^{\dagger}\Phi_2)^2 \notag\\
 & 
+ \lambda_3 (\Phi_1^{\dagger}\Phi_1)(\Phi_2^{\dagger}\Phi_2)  
+\lambda_4 |\Phi_1^\dagger \Phi_2|^2 + \frac{1}{2}\lambda_5 [(\Phi_1^{\dagger}\Phi_2)^2 + \text{h.c.}], \label{pot1}
\end{align}
where $m_3^2$ and $\lambda_5$ are generally complex, but we assume them to be real for simplicity. 
It is useful to rewrite the soft-breaking $Z_2$ parameter $m_3^2$ through $M^2$~\cite{KOSY} as follows:
\begin{align}
M^2 = \frac{m_3^2}{s_\beta c_\beta}, 
\end{align}
where $s_\beta = \sin\beta$ and $c_\beta = \cos\beta$. 
In the following, we use the shorthand notations  $s_X^{} = \sin X$ and $c_X^{} = \cos X$ for an arbitrary angle $X$. 

The tadpole conditions for $h_1$ and $h_2$ fields,  assuming $v_1\neq 0$ and $v_2 \neq 0$, are given by:
\begin{align}
m_1^2 + \frac{1}{2}v^2\left(\lambda_1 c_\beta^2 + \lambda_{345}s_\beta^2\right) -M^2 s_\beta^2= 0, \\
m_2^2 + \frac{1}{2}v^2\left(\lambda_2 s_\beta^2 + \lambda_{345}c_\beta^2\right) -M^2 c_\beta^2= 0,
\end{align}
where $\lambda_{345}=\lambda_3+\lambda_4+\lambda_5$. 
The mass matrices for the charged states 
$M_\pm^2$ in the basis of ($\omega_1^\pm$,$\omega_2^\pm$)
and 
the CP-odd scalar states $M_{\text{odd}}^2$ in the basis of ($z_1$,$z_2$) are diagonalized as
\begin{align}
R^T(\beta)M_{\pm}^2 R(\beta) = \text{diag}(0,m_{H^\pm}^2),\quad
R^T(\beta)M_{\text{odd}}^2 R(\beta) = \text{diag}(0,m_A^2), 
\end{align} 
where $m_{H^\pm}^2$ and $m_{A}^2$ are the squared masses of $H^\pm$ and $A$:
\begin{align}
m_{H^\pm}^2 &= M^2-\frac{v^2}{2}(\lambda_4+\lambda_5), \quad
m_{A}^2     = M^2-v^2\lambda_5. 
\end{align}
The massless states correspond to the modes $G^\pm$ and $G^0$. 
The mass matrix for the CP-even scalar states is also calculated in the basis of ($h_1',h_2'$) as 
\begin{align}
M_{\text{even}}^2 = \begin{pmatrix}
(M_{\text{even}})_{11}^2 & (M_{\text{even}})_{12}^2 \\
(M_{\text{even}})_{12}^2 & (M_{\text{even}})_{22}^2
\end{pmatrix},  
\end{align}
where each of matrix elements is expressed by 
\begin{align}
(M_{\text{even}})_{11}^2 & = v^2(\lambda_1c_\beta^4 + \lambda_2s_\beta^4 + 2\lambda_{345}c_\beta^2s_{\beta}^2), \\
(M_{\text{even}})_{22}^2 & = \left(1+\frac{\xi}{3}\right)[M^2+v^2(\lambda_1+\lambda_2 - 2\lambda_{345})s_{\beta}^2c_{\beta}^2], \\
(M_{\text{even}})_{12}^2 & = v^2\left(1+\frac{\xi}{6}\right)\left[-\lambda_1 c_\beta^2+\lambda_2 s_\beta^2+c_{2\beta}\lambda_{345}\right]s_{\beta}c_\beta.
\end{align}
This matrix can be diagonalized by the rotation  $R(\theta)$  introduced in Eq.~(\ref{theta}) as 
\begin{align}
&m_h^2=c^2_\theta (M_{\text{even}})_{11}^2 + s^2_\theta (M_{\text{even}})_{22}^2 + 2s_\theta c_\theta (M_{\text{even}})_{12}^2,\\
&m_H^2=s^2_\theta (M_{\text{even}})_{11}^2+c^2_\theta (M_{\text{even}})_{22}^2- 2s_\theta c_\theta (M_{\text{even}})_{12}^2,\\
&\tan 2\theta = \frac{2(M_{\text{even}})_{12}^2}{(M_{\text{even}})_{11}^2-(M_{\text{even}})_{22}^2}.
\end{align} 
Now, we can rewrite all the $\lambda_i$ parameters of the potential (\ref{pot1}) in terms of the masses of the physical Higgs bosons and the mixing angle $\theta$ as follows:
\begin{align}
\lambda_1 &= \frac{1}{v^2c^2_\beta}\left[m_h^2c^2_{\beta+\theta}+m_H^2s^2_{\beta+\theta}-M^2s^2_\beta 
+\frac{\xi}{3}s_\beta (m_h^2 c_{\beta+\theta}s_\theta - m_H^2 s_{\beta+\theta}c_\theta  ) \right], \label{lam1_e} \\
\lambda_2 &= \frac{1}{v^2s^2_\beta}\left[m_h^2s^2_{\beta+\theta}+m_H^2c^2_{\beta+\theta}-M^2c^2_\beta 
-\frac{\xi}{3}c_\beta (m_h^2 s_{\beta+\theta}s_\theta + m_H^2 c_{\beta+\theta}c_\theta  ) \right], \\
\lambda_3 &=\frac{1}{v^2}\left[ \frac{2s_{\beta+\theta}c_{\beta+\theta}}{s_{2\beta}}(m_h^2-m_H^2)+ 2m_{H^\pm}^2-M^2
-\frac{\xi}{3s_{2\beta}}(m_h^2 s_\theta c_{2\beta + \theta} -m_H^2 c_\theta s_{2\beta + \theta})  \right], \\
\lambda_4 &= \frac{1}{v^2}(M^2+m_A^2-2m_{H^\pm}^2),  \\
\lambda_5 &= \frac{1}{v^2}(M^2-m_A^2).  \label{lam5_e}
\end{align}

There are in total 9 independent parameters which can be expressed as 
$m_H^{}$, $m_A^{}$, $m_{H^\pm}^{}$, $\cos\theta$, $\tan\beta$, $M^2$, $\xi$ (or $f$), $v$ and $m_h$. The latter two parameters will be fixed in our analysis by requiring  $m_h= 125$ GeV and $v_{\text{SM}}= 246$ GeV. 

\subsubsection{Inert C2HDM}

The Higgs potential is given as in Eq.~(\ref{pot1}) without the $m_3^2$ term.  
Because of the absence of the VEV of $\Phi_2$, 
we have only one tadpole condition for $m_1$, and the $m_2$ parameter will set the scale for the mass  of the inert  Higgs. 
Thus, the mass relations are the following:
\begin{align}
m_{H^\pm}^{2}&=m_{2}^{2}+\frac{\upsilon^{2}}{2}\lambda_{3}, \\
m_{A}^{2}&= {m_{2}}^2+\frac{v^2}{2}  (\lambda_{3}+\lambda_{4}-\lambda_{5}), \\
m_{H}^{2} &=\left(1+\frac{\xi }{3}\right) \left({m_{2}}^2+\frac{v^2}{2}  \lambda_{345}\right), \\
m_h^2 & = \lambda_1 v^2. 
\end{align}
From teh above four relations, the $\lambda_1$, $\lambda_3$, $\lambda_4$ and $\lambda_5$ parameters can be rewritten in terms of 
the four mass parameters and $m_2^2$: 
\begin{align}
\lambda_1 &= \frac{m_h^2}{v^2}, \\
\lambda_3 &= \frac{2}{v^2}(m_{H^\pm}^2-m_2^2), \\
\lambda_4 &= \frac{1}{v^2}\left[m_{A}^2-2m_{H^\pm}^2+m_H^2\left(1-\frac{\xi}{3}\right)\right], \\
\lambda_5 &= \frac{1}{v^2}\left[m_H^2\left(1-\frac{\xi}{3}\right)-m_A^2\right]. 
\end{align}
We note that the $\lambda_2$ parameter is not determined in terms of the Higgs  masses, just like the $m_2^2$ parameter. 

The 8 independent parameters in the potential can be expressed as 
$m_H^{}$, $m_A^{}$, $m_{H^\pm}^{}$, $m_2^2$, $\lambda_2$ and $\xi$ (or $f$), $v$ and $m_h$. 
Similar to the active case, $m_h$ and $v$ will be fixed by 125 GeV and by requiring $v_{\text{SM}}=246$ GeV, respectively.

\section{Unitarity Bounds}

In this section, we discuss the bound from perturbative unitarity in our C2HDM.
We consider all  possible 2-to-2-body bosonic elastic scatterings. 
The procedure to obtain the unitarity bound 
is similar to that in elementary models such as  the SM~\cite{LQT} and  E2HDM~\cite{KKT,Akeroyd,Ginzburg_CPV,Ginzburg,KY}.  
Namely, we compute the $S$-wave amplitude matrix, derive its eigenvalues  $x_i$ and then 
impose the following criterion~\cite{HHG} for each of these: 
\begin{align}
|\text{Re}(x_i)| \leq 1/2. 
\end{align}
 
The most important difference between the unitarity bound in elementary models and that in composite models is that
there is a squared energy dependence in the $S$-wave amplitude for the latter. This is exactly canceled in elementary models among the diagrams with the gauge boson mediation, the Higgs boson mediation and the contact interactions.
In composite models, however, this cancellation does not work, because the sum rule of the Higgs-Gauge-Gauge type couplings is modified from 
that in the elementary ones. 
For example, in the E2HDM, the squared sum of the $hVV$ and $HVV$ ($V=W,Z$) couplings is the same as the squared $h_{\text{SM}}VV$ coupling in the SM, while 
in the C2HDM, that is modified by the factor $(1-\xi)$. The energy dependence of the $S$-wave amplitudes leads to unitarity violation and asks for an Ultra-Violet (UV) completion of the C2HDM. The study of the unitarity bounds in this effective theory therefore gives an indication of the scale at which the onset of other effects of the strong sector become relevant.

\subsection{The $W_L^+ W_L^-\to W_L^+W_L^-$ Reference Process }

In order to clearly show the difference of perturbative unitarity properties in the E2HDM and those in the C2HDM,
let us calculate the elastic scattering of 
the longitudinal component of the $W$ boson scattering, i.e., 
$W_L^+W_L^-\to W_L^+W_L^-$, in the active case.

The contribution from the diagrams without the Higgs bosons
is calculated as  in the SM,
\begin{align}
&\mathcal{M}(W_L^+W_L^- \to W_L^+W_L^-)_{\text{gauge}} \notag\\
&=\frac{s}{2v_{\text{SM}}^2}(1-c_\phi)-\frac{g_Z^2}{4}\left[(2\cos^2\theta_W - 1)(1+c_\phi)-2\tan^2\frac{\phi}{2}\right]
+\mathcal{O}(s^{-1}), 
\end{align}
where $\phi$ is the scattering angle and $s$ is the squared Center-of-Mass (CM) energy. 
The contribution from the  Higgs boson mediation  ($h$ and $H$) is given by: 
\begin{align}
&\mathcal{M}(W_L^+W_L^-\to  W_L^+W_L^-)_{\text{Higgs}}\notag\\
&=-\frac{s}{2v_{\text{SM}}^2}(1-c_\phi)(1-\xi)
-\frac{2}{v_{\text{SM}}^2}(1-\xi)(m_h^2c_\theta^2 + m_H^2s_\theta^2)+\mathcal{O}(s^{-1}).
\end{align}
Thus, in the total amplitude the $s$ dependence appears which vanishes in the limit of $\xi\to 0$: 
\begin{align}
\mathcal{M}(W_L^+W_L^- \to W_L^+W_L^-)_{\text{tot}}&=
\frac{s\xi}{2v_{\text{SM}}^2}(1-c_\phi) 
-\frac{2}{v_{\text{SM}}^2}(m_h^2c_\theta^2 + m_H^2s_\theta^2 )(1-\xi) +\mathcal{O}(g^2,s^{-1}). \label{mtot}
\end{align}  
The $S$-wave amplitude  $a_0$, defined by 
\begin{align}
a_0=\frac{1}{32\pi}\int_{-1}^{1} d\cos\phi\, \mathcal{M} = -\frac{1}{32\pi}\int_{\pi}^{0} d\phi \sin\phi\, \mathcal{M}, 
\end{align}
is calculated for the $W_L^+W_L^- \to  W_L^+W_L^-$ process as 
\begin{align}
a_0(W_L^+W_L^- \to W_L^+W_L^-)&= 
\frac{s}{32\pi v_{\text{SM}}^2}\xi 
-\frac{1}{8\pi v_{\text{SM}}^2}(m_h^2c_\theta^2 + m_H^2s_\theta^2 )(1-\xi)
+\mathcal{O}(g^2,s^{-1}). \label{a0}
\end{align}  
Therefore, $S$-matrix unitarity is broken at a certain energy scale as long as we take $\xi\neq 0$. 

We expect   that exactly the same result as in Eq.~(\ref{a0}), up to ${\cal O}(s^0)$, is obtained
by using the equivalence theorem~\cite{et}, in which 
the $W_L^\pm$ mode is replaced by the Nambu-Goldstone mode $G^\pm$.  Let us check this.
There are three relevant diagrams for the amplitude $(G^+G^- \to G^+G^-)$, i.e., the contact  diagram (denoted by ${\cal M}_c$), and 
the $s$- and $t$-channel diagrams (denoted by ${\cal M}_s$ and ${\cal M}_t$, respectively)  with the $h$ and $H$ exchanges.  
Each of these diagrams is calculated as: 
\begin{align}
\mathcal{M}_{c}(G^+G^- \to G^+G^-)&=\frac{s}{2}(1-c_\phi)(g_{G^\pm G^\pm,G^\mp G^\mp}^{} - g_{G^+ G^-,G^+ G^-})+\lambda_{G^+G^-G^+G^-}, \label{mc_exact} \\
%
\mathcal{M}_s(G^+G^- \to G^+G^-)
&=
-\sum_{\varphi = h,H}\frac{1}{s-m_\varphi^2}\left[\frac{s}{2}(2g_{G^\pm \varphi,G^\mp}^{}-g_{G^+G^-,\varphi}^{} )+\lambda_{G^+G^-\varphi} \right]^2, \label{ms_exact}\\
\mathcal{M}_t(G^+G^- \to G^+G^-)&=-\sum_{\varphi = h,H}\frac{1}{t-m_\varphi^2}\left[\frac{t}{2}(2g_{G^\pm \varphi,G^\mp}^{}-g_{G^+G^-,\varphi}^{} )+\lambda_{G^+G^-\varphi} \right]^2 \label{mt_exact}. 
\end{align}
In the above expressions, we introduced the
scalar trilinear $\lambda_{abc}$ and quartic $\lambda_{abcd}$ couplings from the potential as well as the
scalar trilinear $g_{ab,c}$ and quartic $g_{ab,cd}$ couplings with two derivatives coming from the kinetic Lagrangian. 
They are defined by 
\begin{align}
\lambda_{abcd} \equiv   -\frac{\partial^4 V}{\partial a\partial b\partial c\partial d}, \quad 
\lambda_{abc}   \equiv   -\frac{\partial^3 V}{\partial a\partial b\partial c} \label{lam34}
\end{align}
and
\begin{align}
g_{ab,cd}^{}         \equiv   \frac{\partial^4 {\cal L}_{\text{kin}}}{\partial (\partial_\mu a)\partial(\partial^\mu b)\,\partial c \,\partial d}, \quad 
g_{ab,c}^{}          \equiv   \frac{\partial^3 {\cal L}_{\text{kin}}}{\partial (\partial_\mu a)\partial(\partial^\mu b)\,\partial c}.  \label{g34}
\end{align}
While $g_{ab,cd}^{}$ and $g_{ab,c}^{}$ are proportional to $\xi/v_{\text{SM}}^2$, the 
$\lambda_{abcd}$ and $\lambda_{abc}$ couplings contain  $\xi^0$ terms plus corrections proportional to  $\xi$. 
These scalar couplings appearing in Eqs.~(\ref{mc_exact})--(\ref{mt_exact}) are given by: 
\begin{align}
\lambda_{G^+G^-G^+G^-} &= -\frac{2}{v_{\text{SM}}^2}\left(1 + \frac{\xi}{3} \right)(m_h^2c_\theta^2 + m_H^2s_\theta^2), \label{lamGGGG1}\\
\lambda_{G^+G^-h} &= -\frac{m_h^2}{v_{\text{SM}}}\left(1+\frac{\xi}{6}\right)c_\theta, \quad
\lambda_{G^+G^-H}=  \frac{m_H^2}{v_{\text{SM}}}\left(1+\frac{\xi}{6}\right)s_\theta, \label{lamGGh1},\\
g_{G^+G^-,G^+G^-}^{} &= -\frac{\xi}{3v_{\text{SM}}^2},\quad 
g_{G^\pm G^\pm,G^\mp G^\mp}^{} = \frac{2\xi}{3v_{\text{SM}}^2},  \label{c1} \\
g_{G^+G^-,h} &= -\frac{2\xi}{3v_{\text{SM}}}c_\theta , \quad
g_{G^\pm h,G^\mp} = \frac{\xi}{3v_{\text{SM}}}c_\theta,       \label{c2} \\
g_{G^+G^-,H} &= \frac{2\xi}{3v_{\text{SM}}}s_\theta , \quad
g_{G^\pm H,G^\mp} = -\frac{\xi}{3v_{\text{SM}}}s_\theta.       \label{c3}
\end{align}
By substituting in Eqs. (\ref{mc_exact})--(\ref{mt_exact}), we find that the total amplitude of the $G^+G^- \to G^+ G^-$ process is exactly the same as that of the $W_L^+W_L^- \to W_L^+ W_L^-$ one  given in Eq.~(\ref{mtot}). 
In the following, we calculate all the other 2-to-2-body scattering channels using the equivalence theorem. 

\subsection{Generic Formulae for the 2-to-2-body (Pseudo)Scalar Boson Scatterings}

\begin{figure}[t]
\includegraphics[width=120mm]{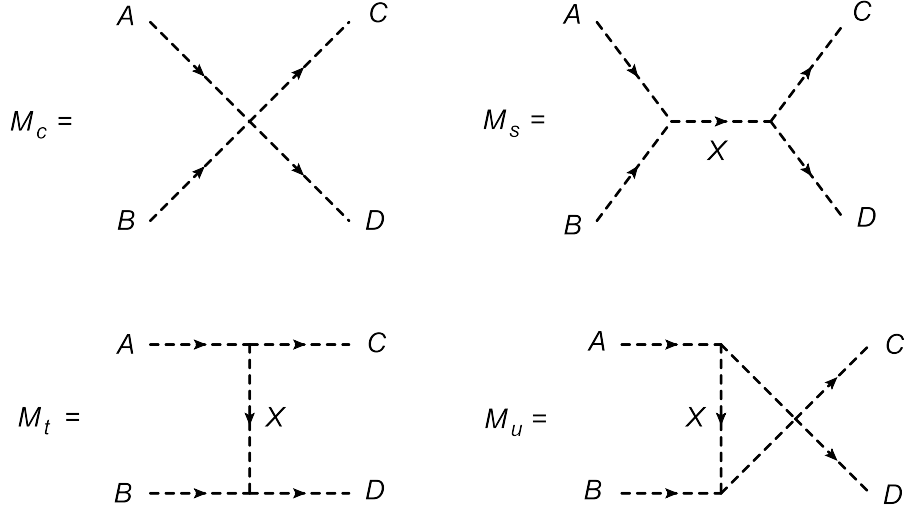} 
\caption{Feynman diagrams for the 2-to-2-body (pseudo)scalar boson scatterings. The arrow with each dashed line shows the momentum flow of each particle. }
\label{diagrams}
\end{figure}

We discuss here the general 2-to-2-body  scattering process
denoted by $A\,B\to C\,D$ with $A,~B,~C$ and $D$
being (pseudo)scalar bosons. 
There are contributions from contact ${\cal M}_c$,  $s$-channel ${\cal M}_s$, $t$-channel ${\cal M}_t$ and $u$-channel ${\cal M}_u$   
diagrams as shown in Fig.~\ref{diagrams}. 
Each of the amplitudes is calculated in the following way:
\begin{align}
{\cal M}_{c}(AB \to CD) &
 = -(g_{AB,CD}^{}\,p_{AB}^{} + g_{CD,AB}^{}\,p_{CD}^{} )\notag\\
&\hspace{-8mm}+ g_{AC,BD}^{}\,p_{AC}^{}+g_{BD,AC}^{}\,p_{BD}^{}+g_{AD,BC}^{}\,p_{AD}^{}+g_{BC,AD}^{}\,p_{BC}^{} + \lambda_{ABCD}^{},   \label{mc}\\
{\cal M}_{s}(AB \to X \to  CD) 
& =   -\frac{1}{s-m_X^2}
\left(g_{XA,B}^{}\, p_{XA}^{} + g_{BX,A}^{}\, p_{BX}^{}-g_{AB,X}^{} \, p_{AB}^{}  +\lambda_{ABX}\right)\notag\\ 
&\quad\quad\quad \times  \left(g_{XC,D}^{}\, p_{XC}^{} + g_{DX,C}^{}\, p_{DX}^{}-g_{CD,X}^{}\, p_{CD}^{}  +\lambda_{CDX} \right),  \label{ms} \\
{\cal M}_{t}(AB \to X \to  CD) 
& =   -\frac{1}{t-m_X^2}
\left(g_{AC,X}^{} \, p_{AC}^{} + g_{XA,C}^{}\, p_{XA}^{} - g_{CX,A}^{}\, p_{CX}^{} +\lambda_{ACX}\right)\notag\\ 
&\quad\quad\quad \times  \left(g_{BD,X}^{}\, p_{BD}^{} - g_{XB,D}^{}\, p_{XB}^{} +g_{DX,B}^{}\, p_{DX}^{}+\lambda_{BDX} \right),  \label{mt}\\
{\cal M}_{u}(AB \to X \to  CD) 
& =   -\frac{1}{u-m_X^2}
\left(g_{AD,X}^{} \, p_{AD}^{} + g_{XA,D}^{}\, p_{XA}^{} - g_{DX,A}^{}\, p_{DX}^{} + \lambda_{ADX}\right)\notag\\ 
&\quad\quad\quad \times  \left(g_{BC,X}^{}\, p_{BC}^{} - g_{XB,C}^{}\, p_{XB}^{} +g_{CX,B}^{}\, p_{CX}^{}+\lambda_{BCX} \right),  \label{mu}
\end{align}
where $p_{ij} = p_i\cdot p_j$. In the above expression, $\lambda_{abc}$ and 
$\lambda_{abcd}$ are defined in Eq.~(\ref{lam34}) while $g_{a,bc}$ and $g_{ab,cd}$ are defined in Eq.~(\ref{g34}). 
The four-momenta of the particles $A$, $B$, $C$ and $D$ are expressed as 
\begin{align}
p_i^{\mu} = (E_i, \, \vec{p}_i), \quad (i=A,~B,~C,~D), 
\end{align}
with $E_i$ and $\vec{p}_i$ being the energy and three-momentum of the particle $i$, respectively. 
In the CM frame, these quantities are expressed by 
\begin{align}
&E_A = \frac{\sqrt{s}}{2}(1+x_A^{}-x_B), \quad E_B = \frac{\sqrt{s}}{2}(1+x_B^{}-x_A), \label{e_a}\\
&E_C = \frac{\sqrt{s}}{2}(1+x_C^{}-x_D), \quad E_D = \frac{\sqrt{s}}{2}(1+x_D^{}-x_C), \label{e_b}\\
&\vec{p}_A = (0,0,p_{\text{in}}), \quad \vec{p}_C = (p_{\text{out}}s_\phi,0,p_{\text{out}}c_\phi), \quad \vec{p}_B = -\vec{p}_A,\quad \vec{p}_D = -\vec{p}_C, \label{e_c}
\end{align}
and 
\begin{align}
&p_{\text{in}} = \frac{\sqrt{s}}{2}\lambda^{1/2}(x_A^{},x_B^{}),\quad
 p_{\text{out}} = \frac{\sqrt{s}}{2}\lambda^{1/2}(x_C^{},x_D^{}), \quad
 x_i = \frac{m_i^2}{s}.      \label{x_i}
\end{align}
The 2-body phase space function $\lambda$ is given by 
\begin{align}
\lambda(x,y) = 1 + x^2 + y^2 -2xy-2x-2y. 
\end{align}
In the massless limit of the external particles, i.e., $x_i\to 0$, 
we obtain the simpler form:
\begin{align}
{\cal M}_{c} &
 \to -\frac{s}{2}\Big[g_{AB,CD}^{} + g_{CD,AB}^{}-\frac{1-c_\phi}{2}(g_{AC,BD}^{}+ g_{BD,AC}^{}) \notag\\
&\quad\quad\quad -\frac{1+c_\phi}{2}(g_{AD,BC}^{} + g_{BC,AD}^{})  \Big] + \lambda_{ABCD}^{}, \label{mc22} \\
{\cal M}_{s}
&  \to   -\frac{s^2}{4(s-m_X^2)}
\left(g_{XA,B}^{}\, + g_{BX,A}^{}\, -g_{AB,X}^{}   +\frac{2}{s}\lambda_{ABX} \right)
\times\left(A \to C,\,B\to D \right), \notag\\
&=  -\frac{1}{2}
\left(g_{XA,B}^{}\, + g_{BX,A}^{}\, -g_{AB,X}^{} \right)\lambda_{CDX} + [(A,B)\leftrightarrow (C,D)] + {\cal O}(s^{-1}), \label{ms22} \\
{\cal M}_{t}
&  \to  -\frac{t^2}{4(t-m_X^2)}
\left(g_{AC,X}^{}  - g_{XA,C}^{} - g_{CX,A}^{} -\frac{2}{t}\lambda_{ACX}\right)
\times\left(A \to B,\,C\to D  \right),  \notag\\
&=  -\frac{1}{2}
\left(g_{AC,X}^{}  - g_{XA,C}^{} - g_{CX,A}^{}\right)\lambda_{BDX} + [(A,C)\leftrightarrow (B,D)] + {\cal O}(s^{-1}), \label{mt22} \\
{\cal M}_{u}
& \to  -\frac{u^2}{4(u-m_X^2)}
\left(g_{AD,X}^{}  - g_{XA,D}^{}  - g_{DX,A}^{} -\frac{2}{u}\lambda_{ADX}\right)
\times\left(A \to B,\,D\to C  \right) \notag\\
&=  -\frac{1}{2}
\left(g_{AD,X}^{}  - g_{XA,D}^{}  - g_{DX,A}^{}\right)\lambda_{BCX} + [(A,D)\leftrightarrow (B,C)] + {\cal O}(s^{-1}). \label{mu22}
\end{align}

From the above expressions, 
it is clear that the $S$-wave amplitude can be classified into three types of contributions up to ${\cal O}(s^0)$:  i.e., 
(i) terms proportional to $s\,\xi$, (ii) terms proportional to $s^0\xi^0$ and 
(iii) terms proportional to $s^0\xi$. 
The contributions (i) and (ii)  come only from the $g_{AB,CD}$ and $\lambda_{ABCD}$ coupling, respectively, in the scalar contact interaction diagram as it is seen in Eq.~(\ref{mc22}). 
The contribution (iii) comes from the cross term $g_{A,BC}^{}\times \lambda_{ABC}^{}$ in the $s$, $t$ and $u$ channel diagrams 
and also from the contact diagram in Eq.~(\ref{mc}). 
When we neglect the contribution (iii), the calculation of the $S$-wave amplitude becomes extremely simple for the following reason. 
In this approximation,
the propagator of (pseudo)scalar bosons 
and the invariant mass term from the product of the momentum $p_{ij}$ do not enter. 
We thus can choose any basis of scalar states. 
In other words, the eigenvalues of the $S$-wave matrix do not depend on the mixing angles $\beta$ and $\theta$ for the scalar bosons\footnote{Even the shift of scalar fields given in Eq.~(\ref{shifts}) 
is not needed in this calculation, because the $\xi$ factor from  the shift provides ${\cal O}(\xi^2 s)$ or ${\cal O}(\xi s^0)$ contributions. }. 
Clearly, the simplest way to calculate the $S$-wave matrix is using the weak eigenbasis and we adopt it to 
calculate the $S$-wave amplitudes for all the scattering states in the next subsection. 

Before calculating all the scattering amplitudes, 
let us consider  another particular process, e.g., $H^+ H^- \to H^+H^-$, again in the active case,
in order to see if the ${\cal O}(\xi s^0)$ term can be relevant. 
Using Eqs.~(\ref{mc})--(\ref{mt}), we obtain the amplitude for the $H^+ H^- \to H^+H^-$ process as follows:
\begin{align}
 {\cal M}& (H^+H^- \to H^+H^-) 
= \frac{ s}{2v_{\text{SM}}^2}\xi(1+c_\phi) -\frac{m_{H^\pm}^2}{v_{\text{SM}}^2}\xi \left(\frac{2}{3}+4c_\phi \right) +\lambda_{H^+H^-H^+H^-}  \notag\\
&-\sum_{\varphi=h,H}\lambda_{H^+H^-\varphi}^2\left(\frac{1}{s-m_{\varphi}^2} + \frac{1}{t-m_{\varphi}^2}\right) \label{mc_exact2}\\
& 
= \frac{ s}{2v_{\text{SM}}^2}\xi(1+c_\phi) -\frac{m_{H^\pm}^2}{v_{\text{SM}}^2}\xi \left(\frac{2}{3}+4c_\phi \right) +\lambda_{H^+H^-H^+H^-} + {\cal O}(s^{-1}),   \label{hphm}
\end{align}
and the  couplings relevant to this process are given by: 
\begin{align}
\lambda_{H^+H^-H^+H^-} &=  \frac{2}{v_{\text{SM}}^2}\left(1 - \frac{\xi}{3} \right)\left[4\cot^22\beta\, M^2
-\left(c_\theta + 2\cot2\beta s_\theta\right)^2\, m_h^2
-\left(s_\theta - 2\cot2\beta c_\theta\right)^2\, m_H^2\right]\notag \\
&+\frac{4c_{2\beta}}{3v_{\text{SM}}^2 s_{2\beta}^2}\xi\Big[ (c_\theta s_{2\beta}+2s_\theta c_{2\beta})s_\theta \,m_h^2
+(2c_\theta c_{2\beta}-s_\theta s_{2\beta})c_\theta\, m_H^2 \Big],    \label{4hp} \\
\lambda_{H^+H^- h} &= \frac{1}{v_{\text{SM}}^{}}\left[\frac{2s_{2\beta+\theta}}{s_{2\beta}}M^2 - (c_\theta + 2s_\theta \cot2\beta)m_h^2 - 2c_\theta m_{H^\pm}^2 \right] \notag\\
&+ \frac{\xi}{v_{\text{SM}}^{}}\left[-\frac{c_\theta}{3}M^2 +\frac{1}{6}(c_\theta + 4s_\theta \cot2\beta)m_h^2 +\frac{c_\theta}{3} m_{H^\pm}^2 \right], \label{3hp1} \\
\lambda_{H^+H^- H} &= \frac{1}{v_{\text{SM}}^{}}\left[\frac{2c_{2\beta+\theta}}{s_{2\beta}}M^2 + (s_\theta - 2 c_\theta \cot2\beta)m_H^2 + 2s_\theta m_{H^\pm}^2 \right] \notag\\
&+ \frac{\xi}{v_{\text{SM}}^{}}\left[\frac{s_\theta}{3}M^2 +\frac{1}{6}(-s_\theta + 4c_\theta \cot2\beta)m_H^2 - \frac{s_\theta}{3} m_{H^\pm}^2 \right], \label{3hp2} 
\end{align}
and 
\begin{align}
g_{H^+H^-,H^+H^-}^{} &= -\frac{\xi}{3v_{\text{SM}}^2},\quad 
g_{H^\pm H^\pm,H^\mp H^\mp}^{} = \frac{2\xi}{3v_{\text{SM}}^2},  \\
g_{H^+H^-,h} &= g_{H^\pm h,H^\mp} = g_{H^+H^-,H} = g_{H^\pm H,H^\mp} = 0.       \label{3h}
\end{align}
Notice that the contribution from the $s$- and $t$-channels with  $h$ and $H$ mediation only give the ${\cal O}(s^{-1})$ term in the $H^+H^- \to H^+H^-$ amplitude
due to the absence of the trilinear $g_{a,bc}$ couplings as shown in Eq.~(\ref{3h}).  
In contrast, the $s$- and $t$-channel contributions to the  $G^+G^- \to G^+G^-$ amplitude do
give ${\cal O}(\xi s^0)$ terms. 
In addition, the second term in Eq.~(\ref{hphm}) comes from the mass dependence of the product of the four momenta (see Eqs.~(\ref{e_a})--(\ref{x_i})) 
in the contact interaction diagram.

\begin{figure}[t]
\begin{center}
\includegraphics[width=55mm]{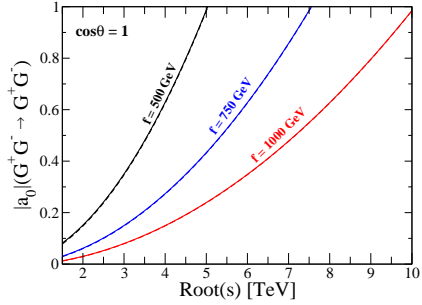}
\caption{$S$-wave amplitude for the $G^+G^-\to G^+G^-$ process as a function of $\sqrt{s}$ in the case of $\cos\theta = 1$
and  $f=$ 500 GeV (black), 750 GeV (blue), 1000 GeV (red).  
The solid (dashed) curves are the result with (without) ${\cal O}(\xi s^0)$ terms. 
}
\label{fig:a0ww}
\end{center}
\end{figure}

\begin{figure}[t]
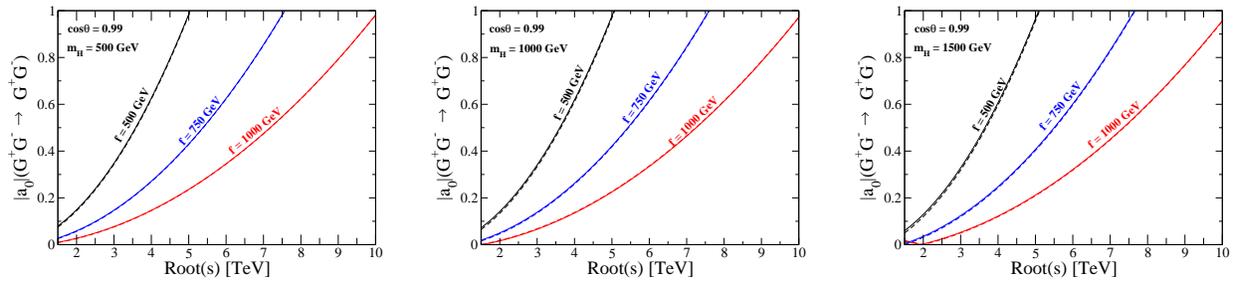

\begin{center}
\includegraphics[width=50mm]{a0_WW_mH500.eps}\hspace{5mm}
\includegraphics[width=50mm]{a0_WW_mH1000.eps}\hspace{5mm}
\includegraphics[width=50mm]{a0_WW_mH1500.eps}
\caption{{$S$-wave amplitude for the $G^+G^-\to G^+G^-$ process as a function of $\sqrt{s}$ in the case of $\cos\theta = 0.99$ and  $f=$ 500 GeV (black), 750 GeV (blue), 1000 GeV (red).  
The solid (dashed) curves are the result with (without) ${\cal O}(\xi s^0)$ terms. 
The left, center and right panels show the result for $m_H^{}=500$, 1000 and 1500 GeV, respectively. }
}
\label{fig:a0ww2}
\end{center}
\end{figure}

\begin{figure}[t]
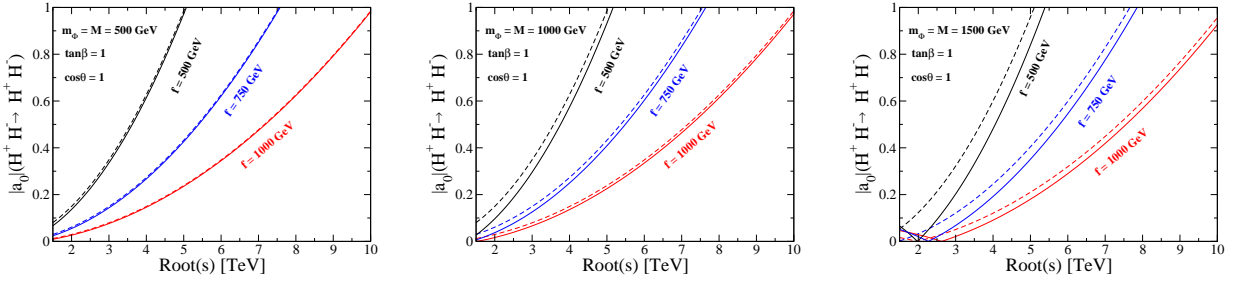

\begin{center}
\includegraphics[width=50mm]{a0_HpHm_500.eps}\hspace{5mm}
\includegraphics[width=50mm]{a0_HpHm_1000.eps}\hspace{5mm}
\includegraphics[width=50mm]{a0_HpHm_1500.eps}
\caption{{$S$-wave amplitude for the $H^+H^-\to H^+H^-$ process as a function of $\sqrt{s}$ in the case of $\cos\theta=1$, $\tan\beta = 1$ and
 $f=500$ (black), 750 (blue) and 1000 GeV (red). 
The solid (dashed) curves are the results with (without) ${\cal O}(\xi s^0)$ terms. 
The left, center and right panels show the results for $m_\Phi^{}(=m_A^{}=m_H^{}=m_{H^\pm}) = M=500$, 1000 and 1500 GeV, respectively. }
}
\label{fig:a0hphm1}
\end{center}
\end{figure}

Let us now show some numerical results for the $S$-wave amplitudes of the $G^+G^- \to G^+G^-$ and $H^+H^- \to H^+H^-$ scatterings. 
{First, we show the results by neglecting the ${\cal O}(s^{-1})$ terms in order to see fully the effect of the ${\cal O}(\xi s^0)$ contributions.
In Figs.~\ref{fig:a0ww} and \ref{fig:a0ww2}, we plot the absolute value of $a_0(G^+G^-\to G^+G^-)$ 
as a function of $\sqrt{s}$ in the case of $\cos\theta = 1$ and $\cos\theta=0.99$, respectively. 
In the both figures, the solid (dashed) curves show the case with (without) ${\cal O}(\xi s^0)$ contributions
and the scale $f$ is taken to be 500 GeV (black), 750 GeV (blue) and 1000 GeV (red). 
In Fig.~\ref{fig:a0ww2}, we take $m_H=$500, 1000 and 1500 GeV in the left, center and right panels, respectively.
As expected, the $S$-wave amplitude grows as $\sqrt{s}$ increases because of the ${\cal O}$($\xi\,s$) terms, 
so that the unitarity constraint will give an upper limit on $\sqrt{s}$ for a given set of the parameters with $\xi\neq 0$. 
We see that the difference between the solid and dashed curves for each fixed value of $f$ is negligibly small for $\cos\theta=1$ because 
the difference only comes from the $m_h^2$ term, as shown in Eq.~(\ref{a0}), whereas the $m_H$ dependence vanishes.
For the case with $\cos\theta=0.99$, a slightly larger difference appears, especially for a larger value of $m_H^{}$, as expected from Eq.~(\ref{a0}). 
Although a further larger difference is expected to appear as $\theta$ increases for a fixed value of $m_H$, 
such a scenario is disfavored by the current LHC data~\cite{LHC_ATLAS2,LHC_CMS2}, 
which causes a large deviation in the $hVV$ coupling from the SM value.  
So, in summary, we can safely neglect the ${\cal O}(\xi s^0)$ contributions in the $S$-wave amplitude for the $G^+G^- \to G^+G^-$ process.
}

In Fig.~\ref{fig:a0hphm1}, we show the $S$-wave amplitude for the $H^+H^-\to H^+H^-$ scattering as a function of $\sqrt{s}$ in the case of 
$\cos\theta = 1$, $\tan\beta = 1$ and $M = m_\Phi^{}(= m_A^{}=m_H^{}=m_{H^\pm})$.  
Similarly to Fig.~\ref{fig:a0ww}, the solid and dashed curves  show the cases with and without ${\cal O}(\xi s^0)$ terms, respectively. 
Because of the $m_{H^\pm}^2\,\xi$ term in Eq.~(\ref{hphm}), 
the difference between these two cases becomes larger when we take  larger values of $m_\Phi^{}$ and small $f$. 

\begin{figure}[t]
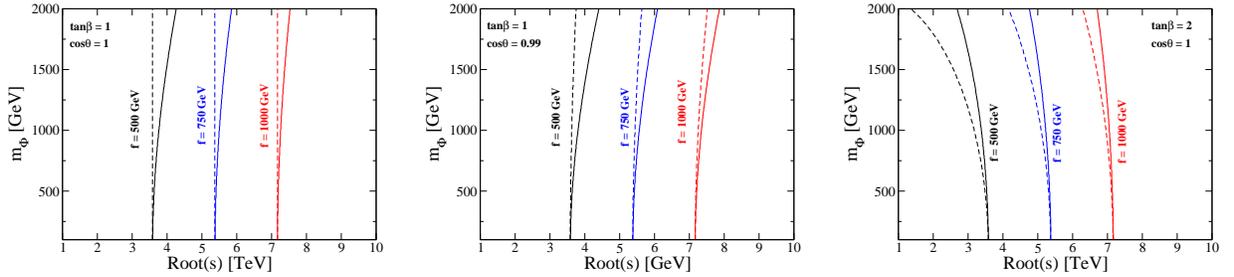

\begin{center}
\includegraphics[width=50mm]{bound_HpHm_ct1.eps} \hspace{3mm}
\includegraphics[width=50mm]{bound_HpHm_ct099.eps} \hspace{3mm}
\includegraphics[width=50mm]{bound_HpHm_ct1_tanb2.eps} 
\caption{Unitarity bound on the ($\sqrt{s}$, $m_\Phi^{}$) plane from the requirement of $|a_0(H^+H^- \to H^+H^-)|<1/2$ in the case of $M=m_\Phi^{}(=m_A^{}=m_H^{}=m_{H^\pm})$. 
In the left, center and right panels, we take $(\cos\theta,\tan\beta)=(1,1)$, (0.99,1) and (1,2), respectively. 
The solid (dashed) curves are the result with (without) ${\cal O}(\xi s^0)$ terms. 
}
\label{fig:const_hphm}
\end{center}
\end{figure}

In Fig.~\ref{fig:const_hphm}, 
we show the constraints on the ($\sqrt{s}$, $m_\Phi^{}$) plane by the requirement that 
the magnitude of $a_0(H^+H^- \to H^+H^-)$ does not exceed 1/2. 
When we include the ${\cal O}(\xi s^0)$ contribution, the constraints become slightly weaker 
because of the destructive contributions between the $\xi\,s$ and $\xi\,m_{H^\pm}^2$ term.

From the above results shown in Figs.~\ref{fig:a0ww}-\ref{fig:const_hphm},  
we can conclude that 
the ${\cal O}(\xi s^0)$ contributions are not so important as long as we consider the case $m_{\Phi}^{}\lesssim 1000$ GeV. 
Since we expect the same holds true for the other 2-to-2-body scalar scattering amplitudes entering the $S$-wave amplitude matrix, we will in the following neglect the ${\cal O}(\xi s^0)$ terms and we will focus our analysis on the region  $m_{\Phi}^{}\lesssim 1000$ GeV where this approximation is safe. Notice also  that this is the region of the C2HDM parameters 
that we are interested in for the phenomenology at the LHC.

\begin{figure}[t]
\begin{center}
\includegraphics[width=50mm]{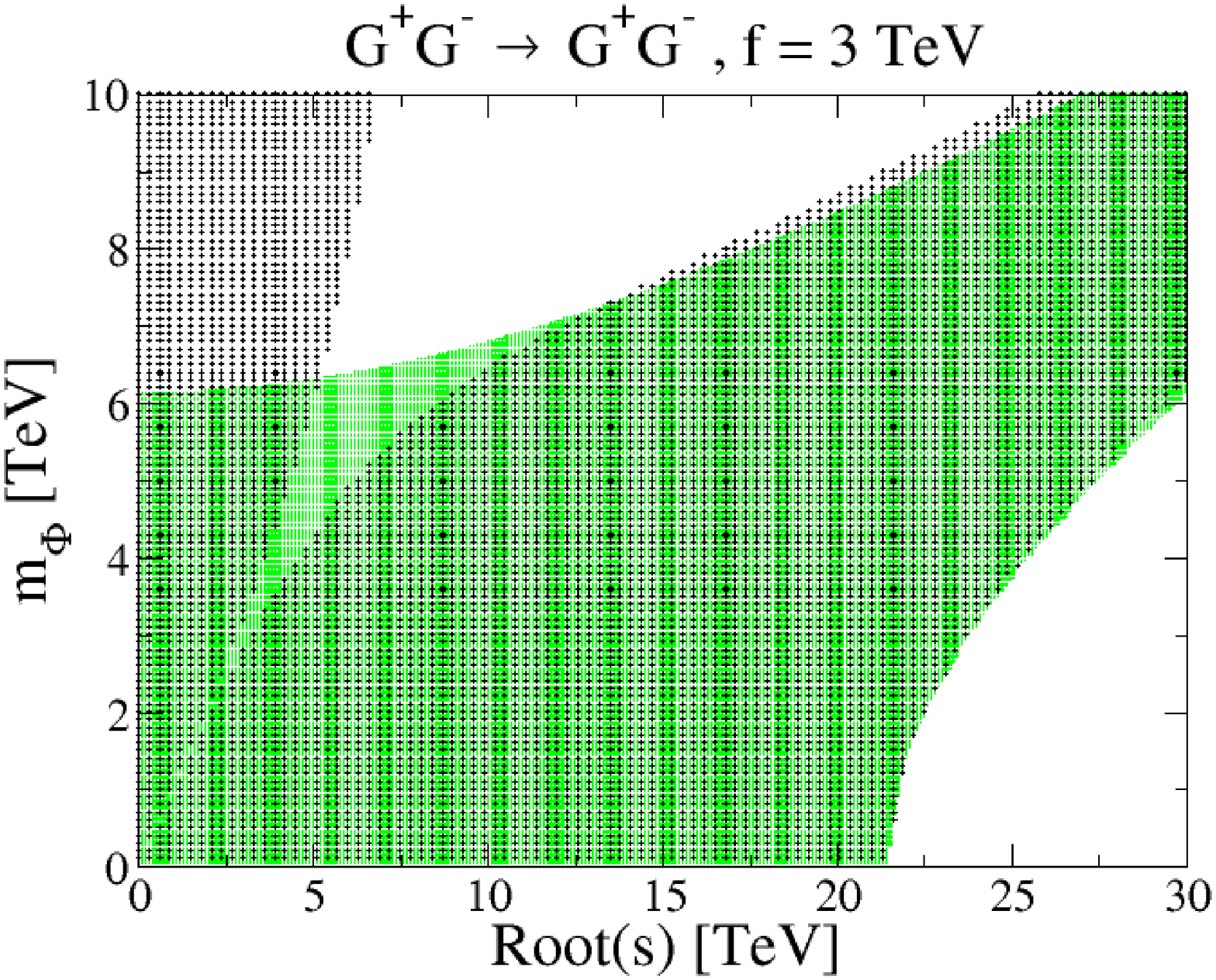} \hspace{3mm}
\includegraphics[width=50mm]{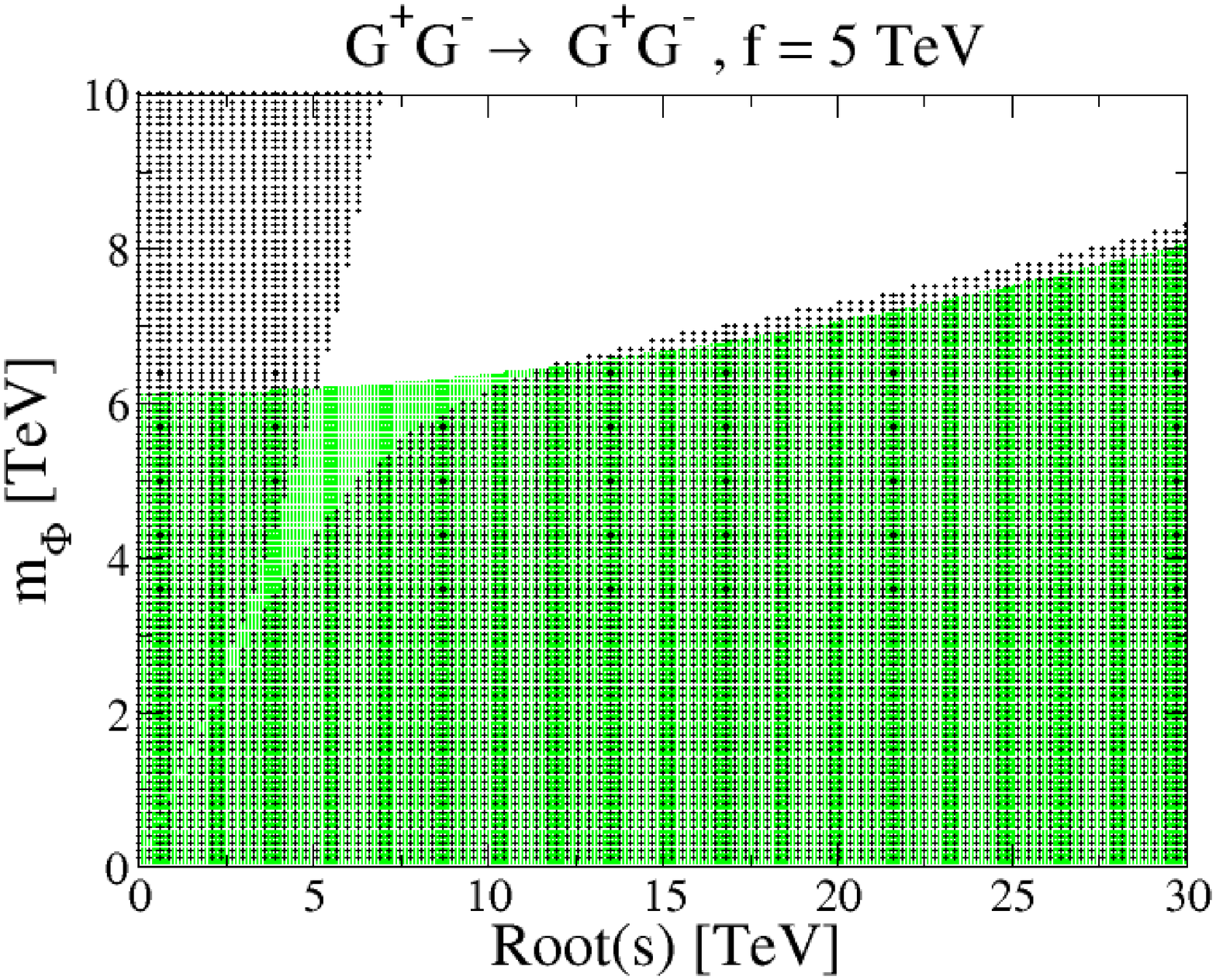} \hspace{3mm}
\includegraphics[width=50mm]{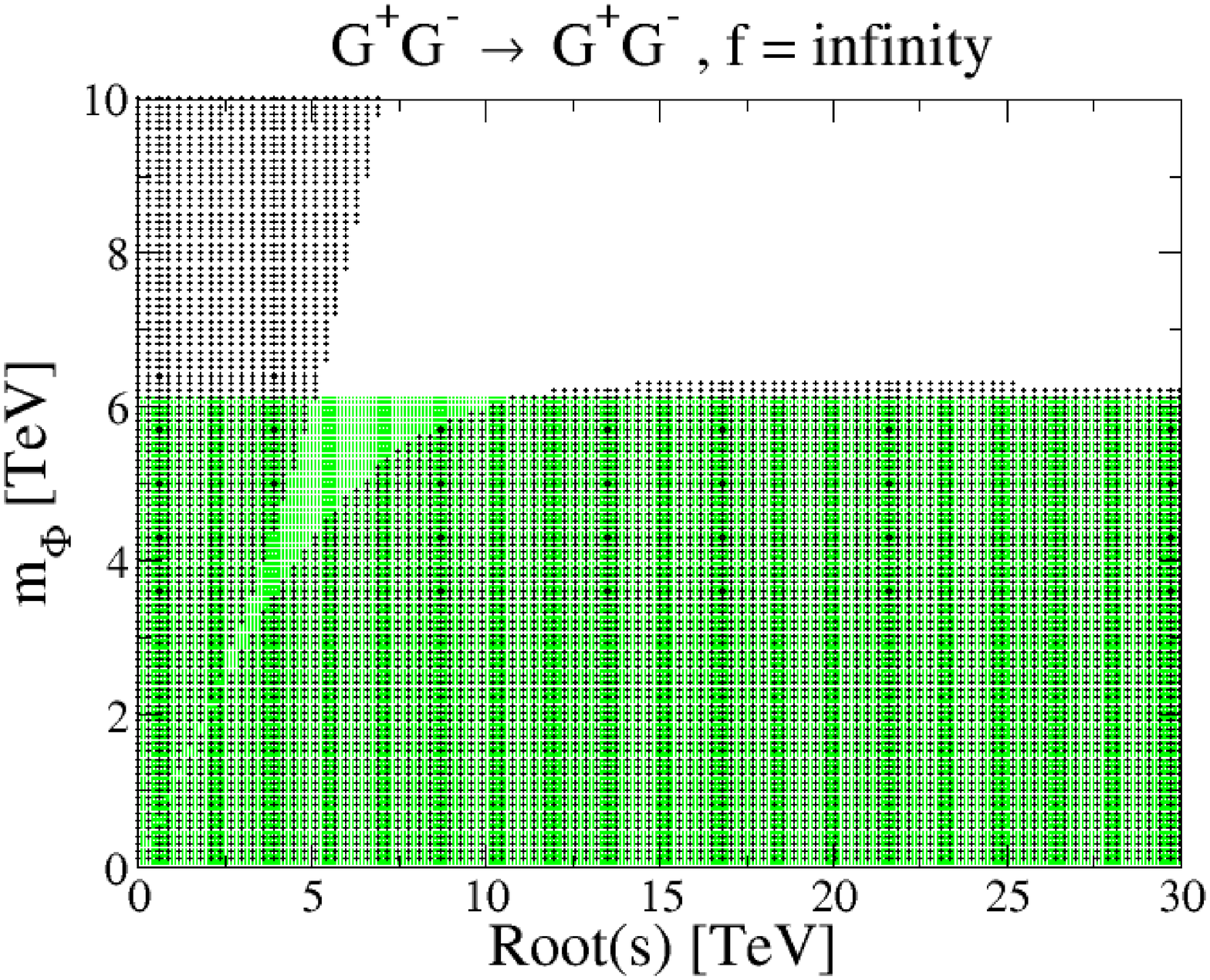} \\  
\includegraphics[width=50mm]{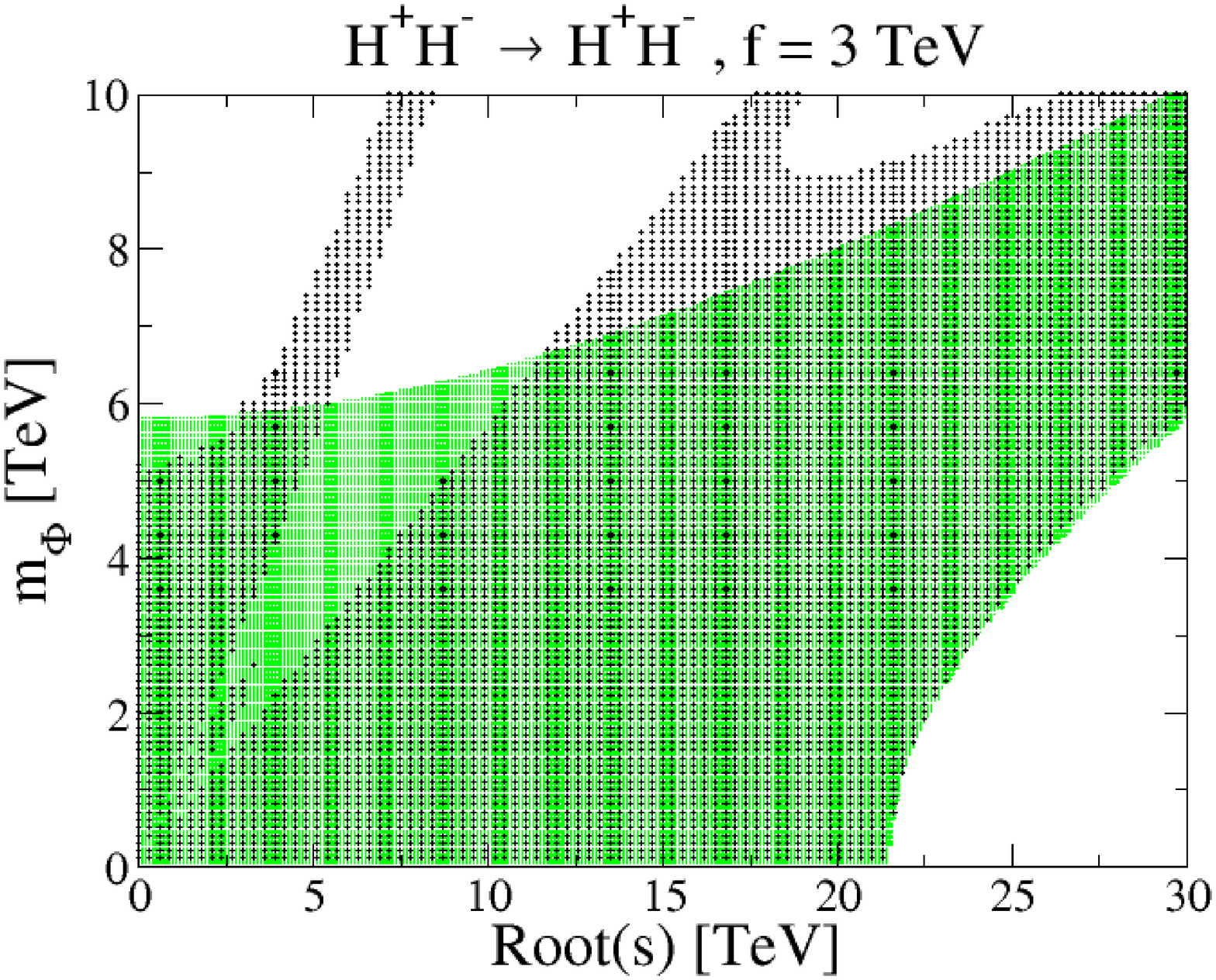} \hspace{3mm}
\includegraphics[width=50mm]{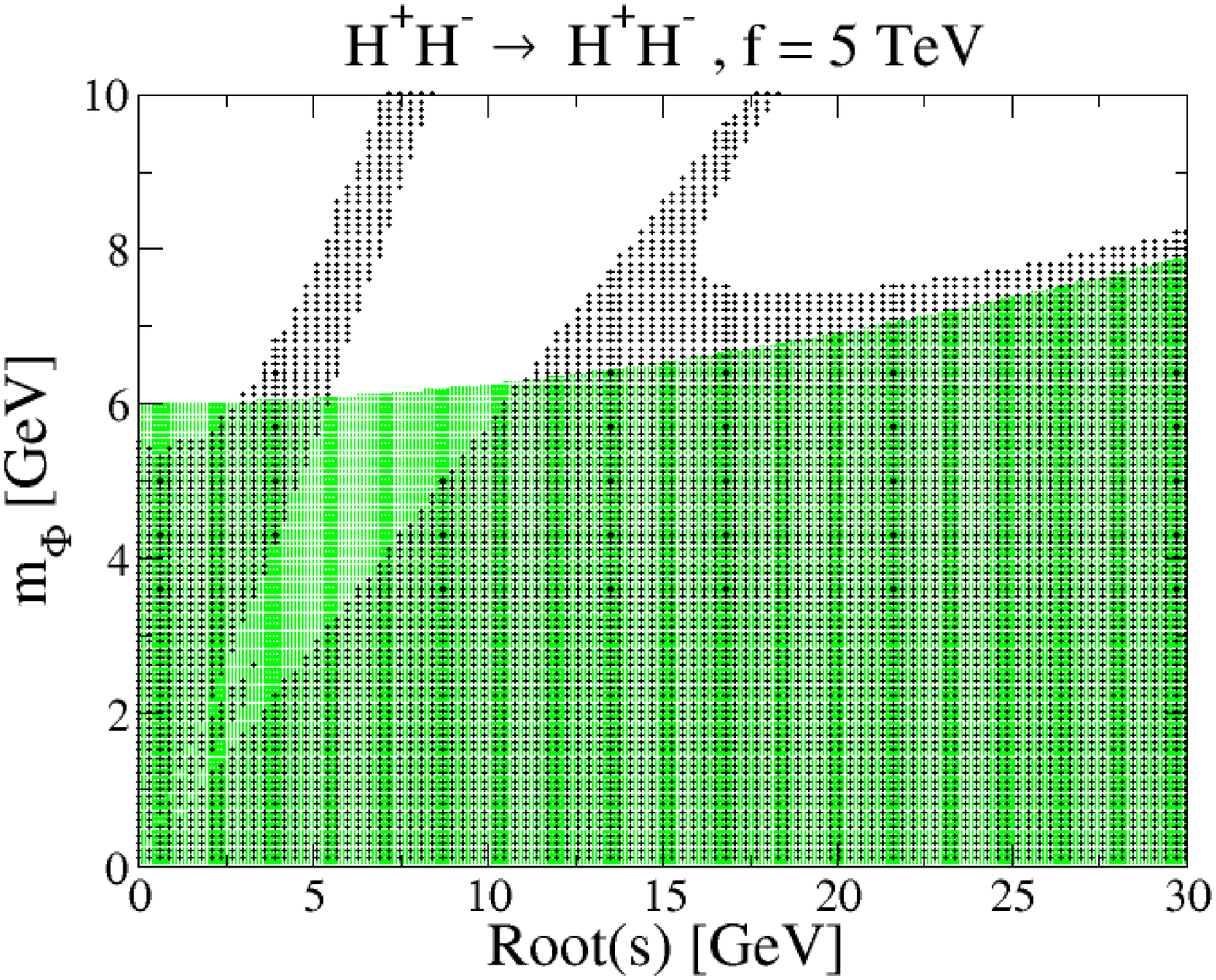} \hspace{3mm}
\includegraphics[width=50mm]{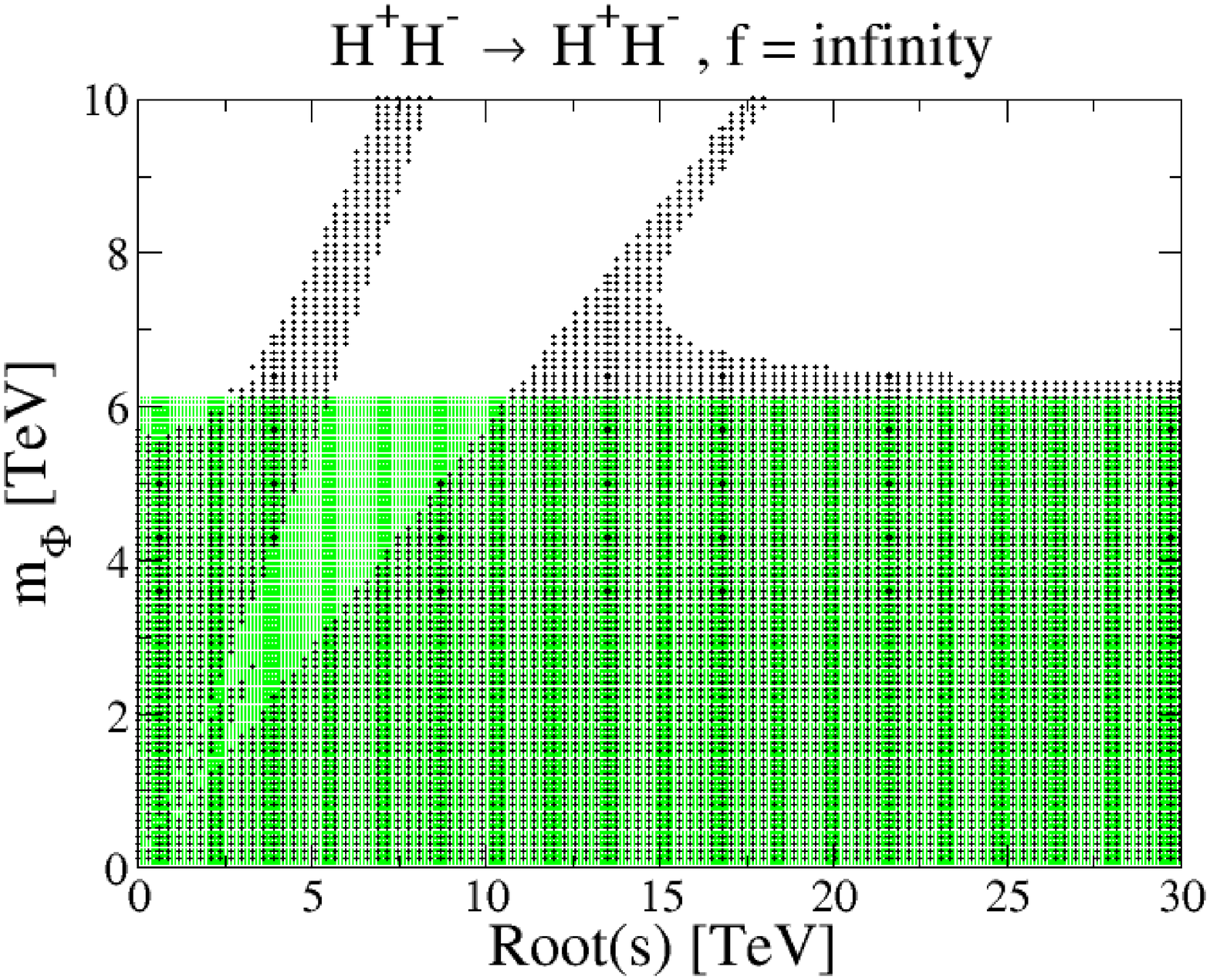}
\caption{Allowed regions from perturbative unitarity in the $(\sqrt s,m_H^{})$ plane from $G^+G^-\to G^+G^-$  (upper panels)
and from $H^+H^-\to H^+H^-$  (lower panels) scattering amplitudes within the C2HDM. 
We take $\cos\theta = 0.99$, $\tan\beta=1$ and $m_\Phi^{}(=m_H=m_A=m_{H^\pm})=M$. 
The grey regions are obtained by using the exact formulae,  the green ones by neglecting ${\cal O}(1/s)$ terms.
The left, center and right panels show the cases with $f=3$ TeV, 5 TeV and infinity (corresponding to the E2HDM).}
\label{shaded}
\end{center}
\end{figure}

Next, we discuss the effect of the ${\cal O}(1/s)$ terms on the $S$-wave amplitudes which were neglected in the above numerical calculations. 
In Fig.~\ref{shaded}, we show the  regions allowed by the unitarity bound using the $G^+G^-\to G^+G^-$ (upper panels) and  
$H^+H^-\to H^+H^-$  (lower panels) scattering amplitudes in the case of $\tan\beta=1$, $\cos\theta=0.99$ and $M=m_\Phi^{}$. 
The black shaded regions show the allowed parameter space using the exact formulae  given in Eqs.~(\ref {mc_exact}), (\ref {ms_exact}), (\ref {mt_exact}) and (\ref{mc_exact2})  
while the green shaded regions do so using 
the approximate formulae given in Eq.~(\ref{a0}) by neglecting the ${\cal O}(1/s)$ terms. 
The value of $f$ is taken to be 3000~GeV, 5000~GeV and infinity (corresponding to the E2HDM) in the left, center and right panels, respectively. 
We see that these two results are in good agreement  for  values of $\sqrt{s}$ large as compared to $m_H^{}$. 
In the complementary region of $ m_H^{}\gtrsim \sqrt{s}$, we find somewhat significant differences between these two results. 
In particular, the region with $\sqrt{s}\simeq m_H^{}$ is excluded if we look at the result using the exact formula, which is 
due to the resonant/divergent effect of the $s$- and $t$-channel diagrams that makes the $S$-wave amplitude quite large. 
Therefore, as long as we consider the phenomenologically interesting case, i.e., the mass of the extra Higgs boson is taken to be ~1 TeV or below and $\sqrt{s}>1$ TeV, 
the differences due to the ${\cal O}(1/s)$ terms are not important either and we can use the approximate formulae to study the unitarity bounds of our model.

Fig.~\ref{shaded} allows also a direct comparison between the C2HDM (left and center panels) and the E2HDM (right panels). 
Given a finite value to $f$, there is an energy scale over which the theory is no longer valid and an UV completion is required 
(for example for $f=3$ TeV we get $\sqrt{s} \lesssim  20$ TeV).  But, for energies below this cut-off, 
the bound on the mass of the extra Higgs boson is less stringent than the one in the E2HDM.  This is due to a partial cancellation between the term growing 
with $s$ and the one proportional to $m_H^2$ 
in the scattering amplitudes here considered 
(or to a squared Higgs mass in general in all other channels). 
This property will be confirmed by the forthcoming  analysis of the unitarity bounds via the complete $S$-wave amplitude matrix.

Furthermore, always in Fig.~\ref{shaded}, one may notice that (e.g., in the left two panels, for $f=$ 3 TeV, where the effect is most apparent)  the C2HDM remains perturbative for very
large values of $\sqrt s$, if also $m_\Phi$ is taken to be large. On the one hand, this corresponds to a very 
fine-tuned region where perturbativity is achieved  through a strong cancellation between the large scalar mass term
and the contribution growing with energy proportionally to $\xi$, the two thereby compensating each other. On the other hand,
over the same region, there are stronger bounds on $m_\Phi^{}$  emerging from 
the $HH\to Hh$ and/or $Hh \to hh$ channels, especially when $\tan\beta \simeq 1$ and $\sin\theta\simeq 0$. 
This is mainly because a larger value of the  $Hhhh$ and $HHHh$ quartic couplings is obtained as compared to 
the $G^+G^-G^+G^-$  and $H^+H^-H^+H^-$ ones. 
For the same configuration given in Fig.~\ref{shaded}, the above neutral (pseudo)scalar channels give an upper limit on $m_\Phi^{}$ of about 2 TeV for $f=3$ TeV and $\sqrt{s}\simeq 0$, i.e., several TeV less than in the charged (pseudo)scalar scattering cases,
with the `funnel region'  onsetting as $\sqrt s\to\infty$ fading away. In general, 
the upper limit on $m_\Phi$ 
becomes stronger when we combine all the  scattering channels together and impose the constraint from vacuum stability as well, as we will see later on in 
Section~III-D.

Finally, despite not presented explicitly here, 
we confirm that the results of the inert case does not differ substantially from the active case, so we shall adopt the same approximations in both constructions.

\subsection{Diagonalization of the $S$-wave Amplitude Matrix}

In this subsection, we calculate all the 2-to-2-body (pseudo)scalar boson scattering amplitudes
by keeping the ${\cal O}(\xi\,s)$ and ${\cal O}(\xi^0\,s^0)$ contributions only.
In this case, the diagonalization of the $S$-wave amplitude matrix is analytically done as we will explain below. 
({We note that the following discussion is valid in both the active and inert case.})

In the C2HDM there are  14 neutral, 8 singly-charged and 3-doubly charged states as in the E2HDM. 
In the weak eigenbasis introduced in Eq.~(\ref{weakeigen}), the 14 neutral channels are expressed by 
\begin{align}
\omega_i^+ \omega_i^-,~
\frac{z_i z_i}{\sqrt{2}},~
\frac{h_i h_i}{\sqrt{2}},~
h_i z_i,~
h_1h_2,~
z_1z_2,~
h_1z_2,~
h_2z_1,~
\omega_1^+ \omega_2^-,~
\omega_2^+ \omega_1^-\quad (i=1,2). 
\end{align}
The 8 (positive) singly-charged channels are expressed by 
\begin{align}
\omega_i^+ z_i,~
\omega_i^+ h_i,~
\omega_1^+ z_2,~
\omega_2^+ z_1,~\omega_1^+ h_2,~\omega_2^+ h_1\quad (i=1,2).  
\end{align}
The 3 (positive) doubly-charged channels are expressed by 
\begin{align}
\frac{\omega_i^+ \omega_i^+}{\sqrt{2}},~
\omega_1^+\omega_2^+\quad (i=1,2). 
\end{align}
The negative charged states are simply obtained by taking the charge conjugation of the corresponding positive states. 

Although each neutral, singly-charged and doubly-charged state respectively gives a $14\times 14$, $8\times 8$ and $3\times 3$ $S$-wave amplitude matrix, respectively,  
they can all be simplified into a block-diagonal form with  maximally $2\times 2$ sub-matrices
by taking appropriate unitary transformations of the scattering states. 
As discussed in Ref.~\cite{Ginzburg_CPV}, 
such an appropriate basis can be systematically obtained by using the conserved quantum numbers, e.g., 
the hypercharge $Y$, the weak isospin $I$, its third component $I_3$ and the $Z_2$ charge of 2-to-2-body scattering states. 

First of all, by using the $Z_2$ charge, we can separate the 14 neutral channels into 8 $Z_2$-even and 6 $Z_2$-odd channels: 
\begin{align}
&\omega_i^+ \omega_i^-,~
\frac{z_i z_i}{\sqrt{2}},~
\frac{h_i h_i}{\sqrt{2}},~
h_i z_i \quad (i=1,2)~~ [Z_2\text{-}\text{even states}], \\
&h_1h_2,~
z_1z_2,~
h_1z_2,~
h_2z_1,~
\omega_1^+ \omega_2^-,~
\omega_2^+ \omega_1^- ~~ [Z_2\text{-}\text{odd states}]. 
\end{align}
Next, the 8 $Z_2$-even states are further decomposed into the following orthogonal states: 
\begin{align}
\frac{1}{\sqrt{2}}\left(\omega_i^+\omega_i^- +\frac{1}{2}z_iz_i + \frac{1}{2}h_ih_i\right),~
\frac{1}{\sqrt{2}}\left(\omega_i^+\omega_i^- -\frac{1}{2}z_iz_i - \frac{1}{2}h_ih_i\right),~
\frac{1}{2}\left(z_iz_i -h_ih_i\right),~z_ih_i. 
\end{align}
The corresponding $8\times 8$ $S$-wave matrix in the above basis is given by 
\begin{align}
a_0^0(Z_2\text{-}\text{even}) &= \text{diag}({\cal A}_1,{\cal A}_2,{\cal A}_3,{\cal A}_4), 
\end{align}
where 
\begin{align}
&{\cal A}_1 = \frac{s\xi}{2v_{\text{SM}}^2}\begin{pmatrix}
3 & 1 \\
1 & 3
\end{pmatrix}-\begin{pmatrix}
3\lambda_1  & 2\lambda_3+\lambda_4 \\
2\lambda_3+\lambda_4 & 3\lambda_2
\end{pmatrix}, \\
&{\cal A}_2 = \frac{s\xi}{2v_{\text{SM}}^2}\begin{pmatrix}
-1 &  1 \\
1 & -1
\end{pmatrix}-\begin{pmatrix}
\lambda_1  & \lambda_4 \\
\lambda_4 & \lambda_2
\end{pmatrix}, ~~{\cal A}_3 = {\cal A}_4 = \frac{s\xi}{2v_{\text{SM}}^2}\begin{pmatrix}
-1 &  1 \\
1 &  -1
\end{pmatrix}-\begin{pmatrix}
\lambda_1  & \lambda_5 \\
\lambda_5 & \lambda_2
\end{pmatrix}. 
\end{align}
The 6 $Z_2$-odd states are further decomposed into the following orthogonal states: 
\begin{align}
&\frac{1}{\sqrt{2}}\left(-z_1z_2+h_1h_2\right),~
\frac{1}{\sqrt{2}}\left(h_1z_2  + h_2z_1\right),~\notag\\
&\frac{1}{2}\left(-z_1z_2 -h_1h_2 + \omega_1^+ \omega_2^- + \omega_2^+ \omega_1^-\right),~
\frac{1}{2}\left(ih_1z_2 - ih_2z_1 - \omega_1^+ \omega_2^- + \omega_2^+ \omega_1^-\right), \notag\\
&\frac{1}{2}\left(z_1z_2 + h_1h_2 + \omega_1^+ \omega_2^- + \omega_2^+ \omega_1^-\right),~
\frac{1}{2}\left(-ih_1z_2 + ih_2z_1 - \omega_1^+ \omega_2^- + \omega_2^+ \omega_1^-\right). 
\end{align}
The corresponding $6\times 6$ $S$-wave matrix in the above basis is given by 
\begin{align}
&a_0^0(Z_2\text{-}\text{odd}) = 
 \frac{s\xi}{v_{\text{SM}}^2}\text{diag}(-1,-1,-1,1,1,1) \notag\\
&- \text{diag}(\lambda_3+\lambda_4,\,
\lambda_3+\lambda_4,\,
\lambda_3+\lambda_5,\,
\lambda_3-\lambda_5,\,
\lambda_3+2\lambda_4+3\lambda_5,\, 
\lambda_3+2\lambda_4-3\lambda_5
). 
\end{align}

Similarly to the neutral states, we can separate the singly-charged states into  4 $Z_2$-even and 4 $Z_2$-odd states:
\begin{align}
&\omega_i^+ z_i,~\omega_i^+ h_i,~~[Z_2\text{-}\text{even states}], \notag\\
&\omega_1^+ z_2,~
\omega_2^+ z_1,~\omega_1^+ h_2,~\omega_2^+ h_1~~[Z_2\text{-}\text{odd states}].
\end{align}
The 4 $Z_2$-even states are further decomposed into the following orthogonal states: 
\begin{align}
\frac{1}{\sqrt{2}}\left(iz_i\omega_i^+ +h_i\omega_i^+\right),~
\frac{1}{\sqrt{2}}\left(-iz_i\omega_i^+ +h_i\omega_i^+\right), 
\end{align}
and the corresponding $4\times 4$ $S$-wave matrix in the above basis is given by 
\begin{align}
&a_0^\pm(Z_2\text{-}\text{even}) = \text{diag}({\cal A}_3,\,{\cal  A}_2). 
\end{align}
The 4 $Z_2$-odd states are further decomposed into the following orthogonal states: 
\begin{align}
&\frac{1}{2}\left(i\omega_1^+ z_2+i\omega_2^+ z_1 +\omega_1^+ h_2 + \omega_2^+ h_1 \right),~ 
\frac{1}{2}\left(-i\omega_1^+ z_2+i\omega_2^+ z_1 -\omega_1^+ h_2 + \omega_2^+ h_1 \right),\notag\\
&\frac{1}{2}\left(-i\omega_1^+ z_2-i\omega_2^+ z_1 +\omega_1^+ h_2 + \omega_2^+ h_1 \right),~ 
\frac{1}{2}\left(i\omega_1^+ z_2-i\omega_2^+ z_1 -\omega_1^+ h_2 + \omega_2^+ h_1 \right), 
\end{align}
and the corresponding $4\times 4$ $S$-wave matrix in the above basis is given by 
\begin{align}
&a_0^\pm(Z_2\text{-}\text{odd}) = 
\frac{s\xi}{v_{\text{SM}}^2}\text{diag}(-1,\,1,\,-1,\,1)
-\text{diag}(\lambda_3+\lambda_4,\,\lambda_3-\lambda_4,\,\lambda_3+\lambda_5,\,\lambda_3-\lambda_5). 
\end{align}

Finally, the 3 doubly-charged states can be separated into  2 $Z_2$-even ($\omega_i^+\omega_i^+/\sqrt{2}$) and 1 $Z_2$-odd state ($\omega_1^+\omega_2^+$). 
They give 
\begin{align}
&a_0^{\pm\pm}(Z_2\text{-}\text{even}) = {\cal A}_3,\quad
a_0^{\pm\pm}(Z_2\text{-}\text{odd}) = -\frac{s\xi}{v_{\text{SM}}^2} -(\lambda_3+\lambda_4). 
\end{align}

Consequently, the analytic formulae of all the independent eigenvalues are obtained by diagonalizing the $2\times 2$ sub-matrices as 
\begin{align}
16\pi x_1^\pm &= \frac{3}{2}\frac{s\,\xi}{v_{\text{SM}}^{2}} -\frac{3}{2}(\lambda_1+\lambda_2)
\pm \frac{1}{2}  \sqrt{9(\lambda_1-\lambda_2)^2+\left(\frac{s\,\xi}{v_{\text{SM}}^{2}}-4\lambda_3-2\lambda_4\right)^2}, \label{x1} \\
16\pi x_2^\pm &= -\frac{1}{2}\frac{s\,\xi}{v_{\text{SM}}^{2}}-
\frac{1}{2}(\lambda_1+\lambda_2)\pm \frac{1}{2}\sqrt{(\lambda_1-\lambda_2)^2+\left(\frac{\xi\,s}{v_{\text{SM}}^2}- 2\lambda_4\right)^2},\\
16\pi x_3^\pm &= -\frac{1}{2}\frac{s\,\xi}{v_{\text{SM}}^{2}}-\frac{1}{2}(\lambda_1+\lambda_2) \pm\frac{1}{2}
\sqrt{(\lambda_1-\lambda_2)^2+\left(\frac{\xi\,s}{v_{\text{SM}}^2}- 2\lambda_5\right)^2},\\
16\pi x_4^\pm &= \frac{s\,\xi}{v_{\text{SM}}^{2}} - (\lambda_3+2\lambda_4\pm 3\lambda_5),\\
16\pi x_5^\pm &= \pm \frac{s\,\xi}{v_{\text{SM}}^{2}}-(\lambda_3\mp\lambda_5), \\
16\pi x_6^\pm &= \pm \frac{s\,\xi}{v_{\text{SM}}^{2}}-(\lambda_3\mp\lambda_4). \label{x6}
\end{align}
{It is important to mention here that the above eigenvalues can be applied to both the active and inert case, as already 
mentioned.  
However, once we rewrite the $\lambda$ parameters in terms of the physical parameters (such as, e.g., the masses of extra Higgs bosons), then 
we obtain different expressions between the active and  inert cases. 
For this reason, the constraints on the physical parameters induced by the unitarity bound could be different in these two cases even  if we use the same expressions for the eigenvalues given in Eqs.~(\ref{x1})--(\ref{x6}). }

\subsection{Constraints by all Channels}

We now perform the numerical evaluations of the theoretical constraints on the C2HDM parameter space induced by the requirement of  perturbative unitarity using all the eigenvalues given in Eqs. (\ref{x1})--(\ref{x6}). 
In addition to the unitarity constraints, we also impose the vacuum stability condition, where 
we require that the scalar potential is bounded from below in any direction of the scalar field space with a large field value. 
The vacuum stability is guaranteed by satisfying the following inequalities~\cite{VS1,VS2}\footnote{We here note that, in general, 
higher order dimensional terms appear in the scalar potential due to non-linear nature and 
these can change the shape of it for large values of the scalar fields,  especially when $f$ is not very large. 
In  total, eight independent dimension six operators, such as $\sim|\Phi_1|^6$, can be written in addition to the terms given in Eq.~(\ref{pot1}), 
which are proportional to $1/f^2$. In our approach, as we explained in Section~II-C, we assume the same form of the potential as in the E2HDM, so that 
we do not take into account the effect of such  higher order operators on the bound from vacuum stability. 
In fact, the potential breaking the EW symmetry in a generic composite Higgs model is generated by loops so that such terms, despite being 
present and participating in the tree level expansion,  are  not responsible for mass generation and 
for inducing a non-zero VEV of the Higgs fields.}:
\begin{align}
\lambda_1 >0, \quad \lambda_2>0,\quad \sqrt{\lambda_1\lambda_2} + \lambda_3 + \text{MIN}(0,\lambda_4\pm \lambda_5) > 0. 
\end{align}

\begin{figure}[t]
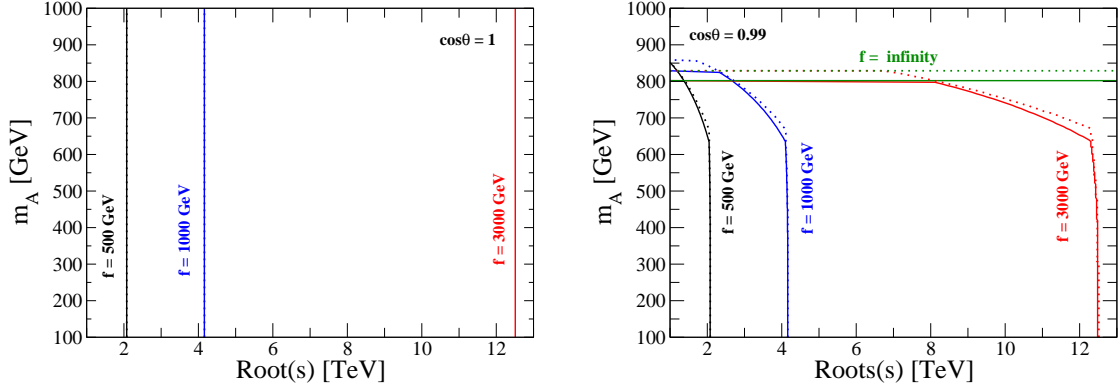

\begin{center}
\includegraphics[width=70mm]{bound_ct1.eps} \hspace{5mm}
\includegraphics[width=70mm]{bound_ct099.eps}
\caption{Constraint on the parameter space of the C2HDM   
from the unitarity and the vacuum stability in the case of $\tan\beta = 1$ and $m_{H^\pm}^{} = m_A^{}$
for several fixed values of $f$. 
The left and right panels show the case with $\cos\theta =1$ and 0.99, respectively. 
The lower left region from each curve is allowed. 
We take the value of $m_{H}^{}$ to be equal to $m_A^{}$ for the solid curves,  
while we scan it within the region of $m_A^{}\pm 500$ GeV for the dashed curves. 
For all the plots, $M$ is scanned. 
}
\label{bound_1}
\end{center}
\end{figure}

We first discuss the constraints for the active C2HDM. 
In Fig.~\ref{bound_1}, we show the allowed parameter regions on the ($\sqrt{s}$, $m_A^{}$) plane for each fixed value of 
$f$, i.e., 500, 1000 and 3000 GeV and infinity (only for the right panel), where $f=\infty$ corresponds to the limit of the E2HDM. 
We take $\cos\theta = 1$ (left) and 0.99 (right). 
In  both panels, $m_{H^\pm}^{}=m_A^{}$ and $\tan\beta=1$ is taken while $M$ is scanned in an enough wide range so as to maximize the allowed parameter region. 
The solid and dashed curves respectively show the case of $m_H^{}=m_A^{}$ and $m_{H}^{}$  scanned within $m_A^{}\pm 500$ GeV. 
We can see from the left panel that there is an upper limit on $\sqrt{s}$,  about 2, 4 and 13 TeV in the case of $f = 500$, 1000 and 3000 GeV, respectively. 
The dependence on $m_A^{}$ for these limits is negligible in the range $m_A^{}\leq 1$ TeV.
If we look at the right panel, we find the limits not only on $\sqrt{s}$ but also on $m_A^{}$, except for the case of $f=\infty$ in which
the limit on $\sqrt{s}$ vanishes as we expect in the E2HDM. 
It is also observed that a bit milder bound on $\sqrt{s}$ and $m_A^{}$ is given 
in the case where we relax the mass degeneracy between $m_A^{}$ and $m_H^{}$ (dashed curves).

\begin{figure}[t]
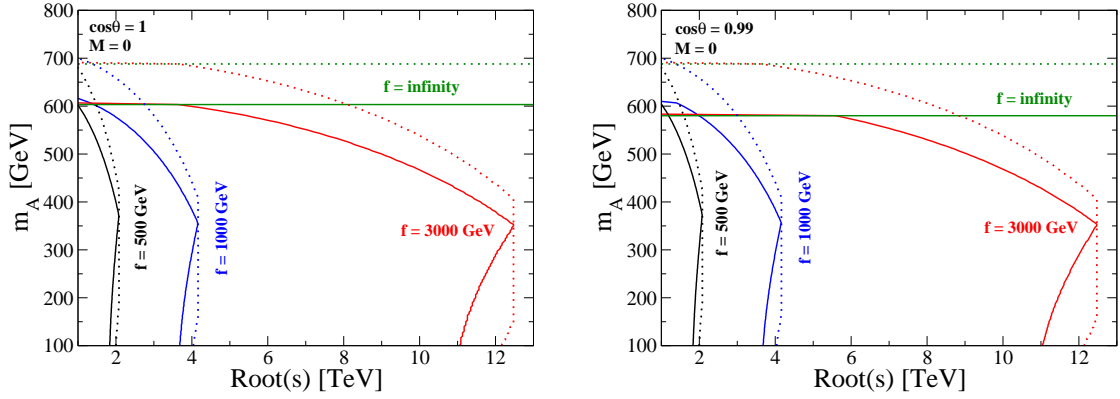

\begin{center}
\includegraphics[width=70mm]{bound_ct1_M0.eps} \hspace{5mm}
\includegraphics[width=70mm]{bound_ct099_M0.eps} 
\caption{Same as Fig.~\ref{bound_1}, but we take  $M=0$. 
}
\label{bound_2}
\end{center}
\end{figure}

In Fig.~\ref{bound_2}, we show the case for $M=0$ by retaining the same configuration used in Fig.~\ref{bound_1}. 
Clearly, a stronger constraint on the ($\sqrt{s}$, $m_A^{}$) plane is provided as compared to the case with scanned $M$. 
According to \cite{so6}, no $M$ term can be generated by the C2HDM potential if the fermions fill the fundamental 6-plet representation of the $SO(6)$ group while 
and a non-zero value of $M$ can be obtained in the C2HDM with traceless symmetric 20-plet fermion representations.  

\begin{figure}[t]
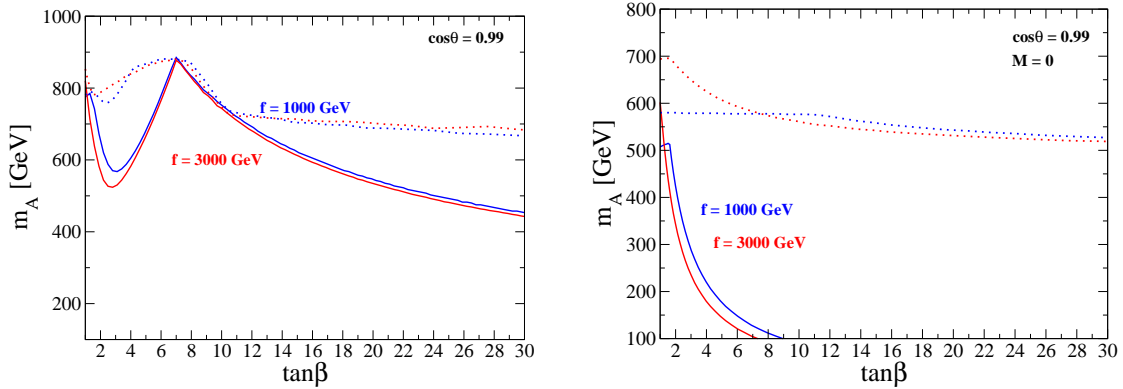

\begin{center}
\includegraphics[width=70mm]{bound_ct099_tanb_p.eps} \hspace{5mm}
\includegraphics[width=70mm]{bound_ct099_tanb_M0_p.eps}
\caption{
Constraint on the parameter space on the ($\tan\beta$, $m_A^{}$) plane  
from the unitarity and the vacuum stability in the case of $\cos\theta = 0.99$, $\sqrt{s}=3000$ GeV and $m_{H^\pm}^{} = m_A^{}$
for $f=1000$ GeV (blue) and $f=3000$ GeV (red). 
The lower left region from each curve is allowed. 
The left panel shows the case with $M$ to be scanned, while the right one does the case with $M = 0$. 
We take the value of $m_{H}^{}$ to be equal to $m_A^{}$ for the solid curves,  
while we scan it within the region of $m_A^{}\pm 500$ GeV for the dashed curves. 
}
\label{bound_3}
\end{center}
\end{figure}

In Fig.~\ref{bound_3}, we show the allowed parameter region on the ($\tan\beta$, $m_A^{}$) plane in the case of $\cos\theta=0.99$, $m_A^{}=m_{H^\pm}$ and $\sqrt{s}=3$ TeV. 
The value of $M$ is scanned in the left panel while it is fixed to be zero in the right panel. 
Similarly to Fig.~\ref{bound_1}, 
the solid and dashed curves show the case of $m_H^{}=m_A^{}$ and $m_{H}^{}$ scanned within $m_A^{}\pm 500$ GeV, respectively.
The case  $f\to \infty$ is almost the same as the case with $f=3000$ GeV. 
We find in the left panel that the case  $\tan\beta \simeq 8$ gives the weakest bound on $m_A^{}$ while for $\tan\beta \gtrsim 10$
the bound gets stronger. 
For the case  $M=0$, the bound is stronger than the case shown in the left panel. 

\begin{figure}[t]
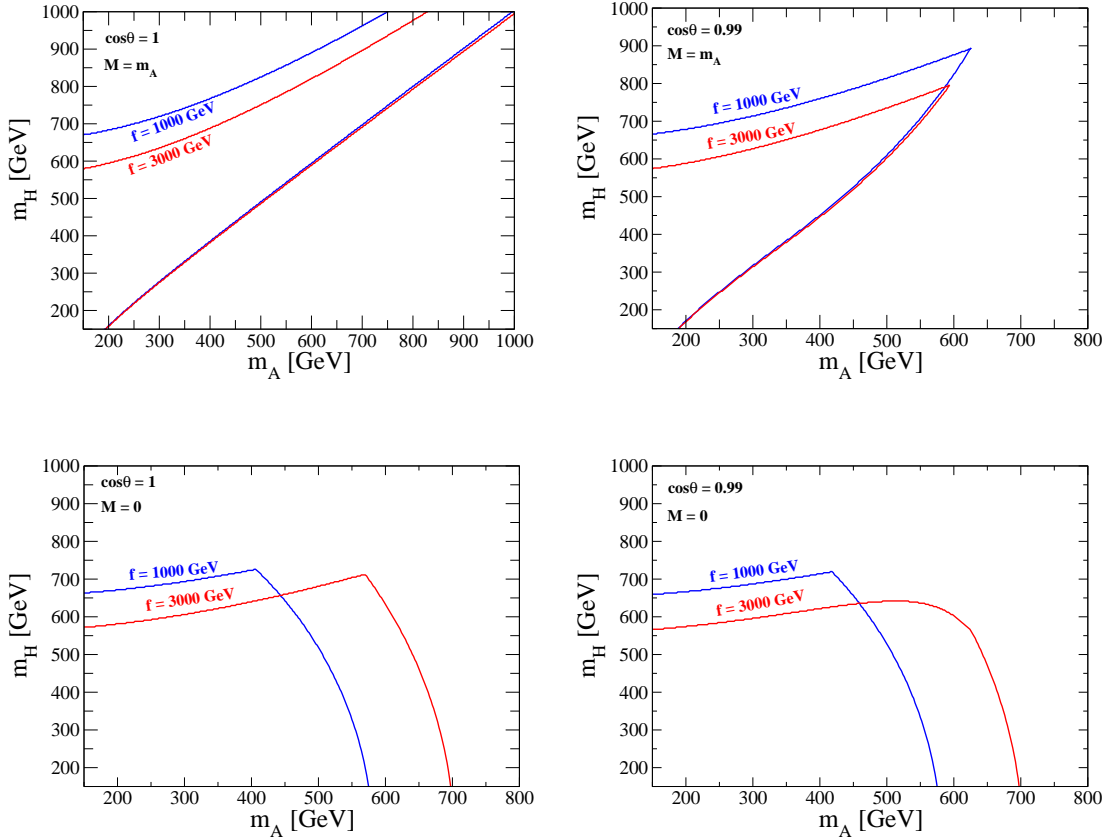

\begin{center}
\includegraphics[width=70mm]{mA-mH_ct1.eps} \hspace{3mm}
\includegraphics[width=70mm]{mA-mH_ct099.eps} \\ \vspace{10mm}
\includegraphics[width=70mm]{mA-mH_M0_ct1.eps} \hspace{3mm}
\includegraphics[width=70mm]{mA-mH_M0_ct099.eps} 
\caption{
Constraint on the parameter space on the ($m_A^{}$, $m_H^{}$) plane by unitarity and vacuum stability in the case with $m_{H^\pm}^{}=m_A^{}$, $\tan\beta=1$ and $\sqrt{s}=3000$ GeV. 
The upper-left, upper-right, lower-left and lower-right panels show the case of 
$(\cos\theta,M)=(1,m_A^{}),~(0.99,m_A^{}),~(1,0)$ and $(0.99,0)$, respectively. 
}

\label{bound_4}
\end{center}
\end{figure}

\begin{figure}[t]
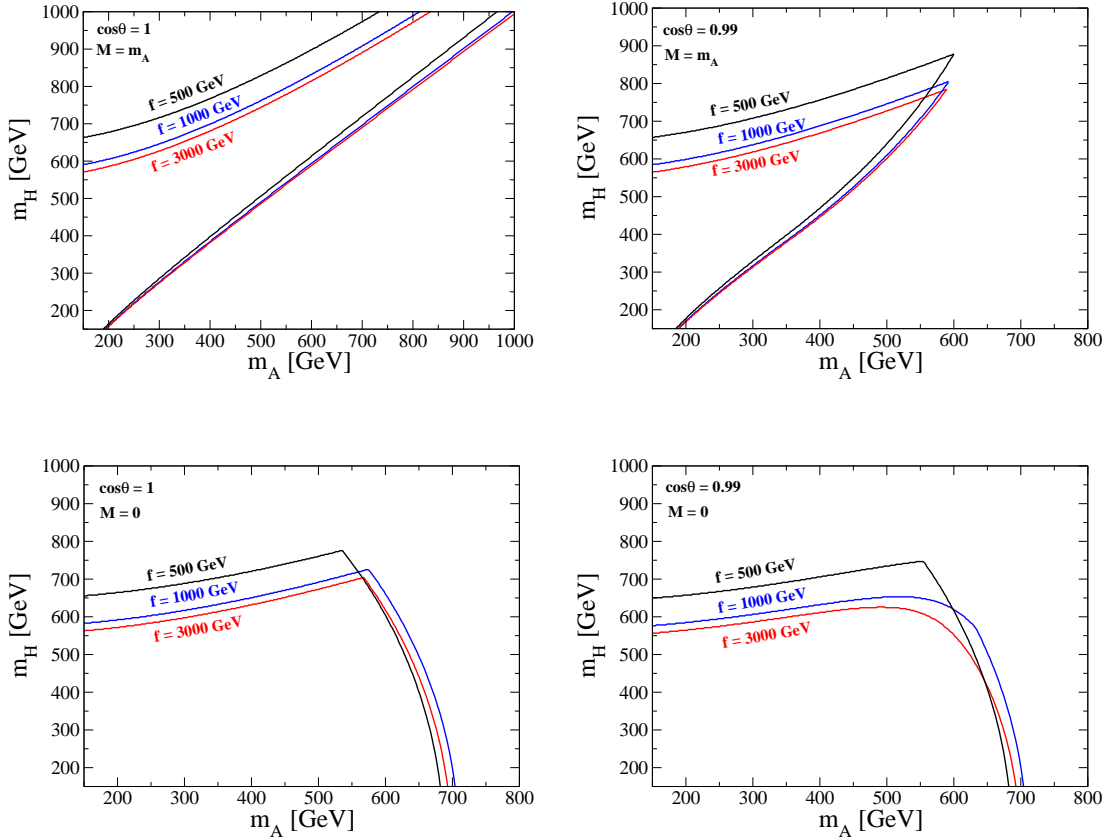

\begin{center}
\includegraphics[width=70mm]{mA-mH_ct1_rs1000.eps} \hspace{3mm}
\includegraphics[width=70mm]{mA-mH_ct099_rs1000.eps} \\ \vspace{10mm}
\includegraphics[width=70mm]{mA-mH_M0_ct1_rs1000.eps} \hspace{3mm}
\includegraphics[width=70mm]{mA-mH_M0_ct099_rs1000.eps} 
\caption{Same as  Fig.~\ref{bound_4} with $\sqrt{s}=1000$ GeV.}
\label{bound_4_roots1TeV}
\end{center}
\end{figure}

In Fig.~\ref{bound_4}, we show the allowed parameter region on the ($m_A^{}$, $m_H^{}$) plane in the case of $m_{H^\pm}^{} = m_A^{}$ and $\sqrt{s}=3000$ GeV.
The values of $(\cos\theta,M)$ are fixed to be $(1,m_A^{})$ for the upper-left,~$(0.99,m_A^{})$  for the upper-right,~(1,0) for the lower-left and $(0.99,0)$ for the lower-right panel. 
For the upper two panels, the region inside the two curves is allowed by unitarity and vacuum stability, where 
the lower (upper) curve is given by the constraint from vacuum stability (unitarity). 
{From the upper two figures, we learn that a too large mass difference between $m_A^{}$ and $m_H^{}$ is not allowed 
by either the unitarity or vacuum stability constraint.  
In addition, if we consider the case for $\cos\theta = 0.99$ (upper-right panel), 
only the region with small masses of $m_A$ and $m_H$, i.e., less than 1 TeV, is allowed  
(we already saw this behavior in the right panel of Fig.~\ref{bound_1}). 
Regarding to the lower panels, we only have an upper bound on $m_H^{}$ and $m_A^{}$ from the unitarity 
requirement, 
whereas the vacuum stability bound does not give a lower limit because  taking $M=0$ 
 render the $\lambda_{1\text{-}5}$ parameters 
 positive (see Eqs.~(\ref{lam1_e})--(\ref{lam5_e})).} Fig.~\ref{bound_4_roots1TeV} is the same as 
 Fig.~\ref{bound_4} with the only difference that we take $\sqrt s=1000$ GeV where  also 
the case $f=500$ GeV is allowed. The distributions here are similar to the case at higher energy, 
with the effect that more parameter space becomes available to the C2HDM with respect to the E2HDM, for smaller $f$ values.

\begin{figure}[t]
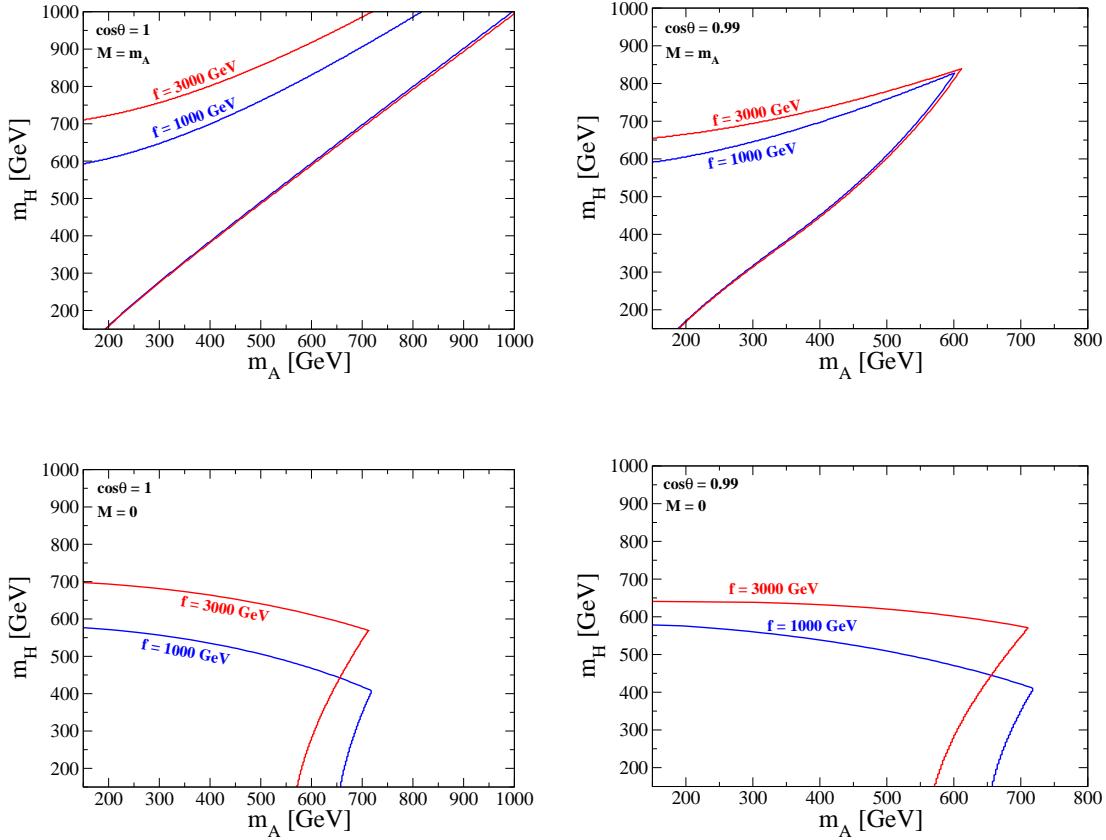

\begin{center}
\includegraphics[width=70mm]{mA-mH_ct1_mHeqmHp.eps} \hspace{3mm}
\includegraphics[width=70mm]{mA-mH_ct099_mHeqmHp.eps} \\ \vspace{10mm}
\includegraphics[width=70mm]{mA-mH_M0_ct1_mHeqmHp.eps} \hspace{3mm}
\includegraphics[width=70mm]{mA-mH_M0_ct099_mHeqmHp.eps} 
\caption{
Constraint on the parameter space on the ($m_A^{}$, $m_H^{}$) plane by unitarity and vacuum stability in the case with $m_{H^\pm}^{}=m_H^{}$, $\tan\beta=1$ and $\sqrt{s}=3000$ GeV. 
The upper-left, upper-right, lower-left and lower-right panels show the case of 
$(\cos\theta,M)=(1,m_H^{}),~(0.99,m_H^{}),~(1,0)$ and $(0.99,0)$, respectively. 
}
\label{bound_4bis}
\end{center}
\end{figure}

Fig.~\ref{bound_4bis} is instead a remake of  Fig.~\ref{bound_4} with $m_{H^\pm}^{} = m_H^{}$. Here, we notice that the distributions of parameter space available in the C2HDM follow an  opposite trend for the same $f$ value. We trace this back to a change of sign in $\lambda_4$, which therefore induces a destructive(constructive) interference (in the
case $m_{H^\pm}^{} = m_H^{}$) when it was instead constructive(destructive) one (in the case $m_{H^\pm}^{} = m_A^{}$).  A similar pattern emerges also for $\sqrt s=1000$ GeV.

\begin{figure}[t]
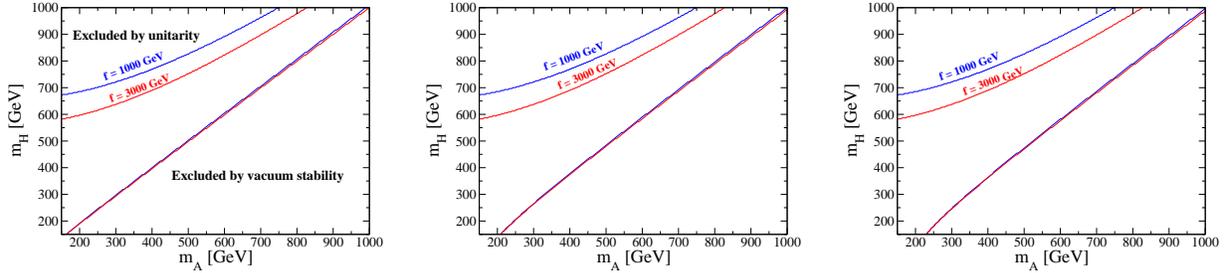

\begin{center}
\includegraphics[width=50mm]{bound_inert.eps} \hspace{3mm} 
\includegraphics[width=50mm]{bound_inert_lam2_2.eps}  \hspace{3mm}
\includegraphics[width=50mm]{bound_inert_lam2_4.eps} 
\caption{
Constraint on the parameter space on the ($m_A^{}$, $m_H^{}$) plane by unitarity and vacuum stability in the inert case for $m_{H^\pm}^{}=m_A^{}=m_2$ and $\sqrt{s}=3000$ GeV. 
We take $\lambda_2 = 0.1$, 2 and 4 in the left, center and right panels, respectively.  
}
\label{bound_5}
\end{center}
\end{figure}

Finally, we briefly discuss the constraints in the inert case. 
In Fig.~\ref{bound_5}, we show the allowed parameter region on the ($m_A^{}$, $m_H^{}$) plane in the case of $m_{H^\pm}^{}=m_A^{}=m_2$ and $\sqrt{s}=3000$ GeV. 
We take $\lambda_2=0.1$ (left), 2 (center) and 4 (right). 
Similarly to the upper panels in Fig.~\ref{bound_4}, the lower and upper curves are respectively determined by the constraints from 
vacuum stability and unitarity while the regions inside the two curves are allowed. 
We see that the vacuum stability bound becomes slightly milder in the case of a larger value of $\lambda_2$ while the unitarity bound is not changed significantly. 
Again, we have here swapped the role of $A$ and $H$ (by requiring $m_{H^\pm}=m_H=m_2$)
as well as lowered $\sqrt s$ to 1000 GeV, like in the case of the active C2HDM, and have found similar patterns to those previously described.

In Fig. 13 we have considered $h$ as the lightest Higgs, 
but a choice of parameters leading to a different mass spectrum is possible. 
For example, we have checked that for $m_H=m_2=100$ GeV the upper limit from unitarity on $m_A~(=m_{H^\pm}^{})$ 
is about 700 GeV. So, a dark matter motivated scenario is available as it is consistent with the unitarity bounds derived in this paper.

\section{Conclusions}

We have studied the bounds from perturbative unitarity as well as vacuum stability in a C2HDM
based on the spontaneous breakdown of a global symmetry $SO(6)\to SO(4)\times SO(2)$ at the compositeness scale $f$. 
We have shown that the ensuing 8 pNGBs can be regarded as two Higgs doublet fields and have derived the kinetic Lagrangian according to 
the CCWZ method.  
We have assumed the same form of the Higgs potential as that in the E2HDM with the softly-broken or exact $Z_2$ symmetry, where 
all the parameters in the potential are taken to be free. 
 
In this construction, we have calculated the $S$-wave amplitude for the elastic 2-to-2-body (pseudo)scalar boson scattering processes. 
We have explicitly shown that the amplitude grows with $\sqrt{s}$ in the 
$W_L^+W_L^-\to W_L^+W_L^-$ (equivalently $G^+G^-\to G^+G^-$) and 
the $H^+H^-\to H^+H^-$ processes as examples, so that unitarity is broken at a certain energy scale depending on the scale $f$.  
We have compared the allowed parameter region from the perturbative unitarity bound in these particular channels 
using the exact formulae and those neglecting ${\cal O}(1/s)$ and/or ${\cal O}(s^0\xi)$ terms. 
We have found that the results using the exact and the approximate formulae
well agree in the region of $\sqrt{s} \gtrsim m_{\Phi}$ ($\Phi=H, A$ or $H^\pm$) and
$m_{\Phi} \lesssim 1$ TeV which is motivated for the LHC phenomenology. 
Therefore, the contribution from ${\cal O}(s^0\xi)$ and ${\cal O}(s^{-1})$ terms can be safely neglected as long as we focus on this 
parameter region, and it allows us to get the explicit analytic expression for the eigenvalues of the $S$-wave amplitude for all the 
possible 2-to-2-body (pseudo)scalar boson scattering, namely, 14 neutral, 8 singly-charged and 3 doubly-charged states. 

We then have numerically demonstrated the allowed parameter space from the unitarity bound 
using all the aforementioned scattering channels and the vacuum stability bound as well. 
In this analysis, we set the
mass of the SM-like Higgs boson $h$ to be 125 GeV, the $hVV$ coupling to be close to the SM value 
(as the discovered Higgs boson is consistent with the SM Higgs boson), 
and taken the masses of the CP-odd and charged Higgs bosons to be degenerate, 
i.e., $m_A=m_{H^\pm}$ (a condition compliant with EW precision data). 
We have also checked how results change by requiring $m_H^{} = m_{H^\pm}$.
We have discovered significant differences of the allowed parameter space 
in the E2HDM and C2HDM that can be exploited in order to separate phenomenologically the two Higgs scenarios. 
The main result that we have found is the following. 
If we take the no-mixing limit between $h$ and $H$, i.e., $\cos\theta=1$ and take the degenerate masses of all the extra Higgs bosons 
then we got the upper limit on $\sqrt{s}$ under the scan of $M^2$, e.g., $\sqrt{s}\lesssim 2$, 4 and 13 TeV for the case of $f=500$, 1000 and 3000 GeV, respectively, as  
we have already seen this behavior in the particular scattering channels $G^+G^-\to G^+G^-$ and $H^+H^-\to H^+H^-$. 
If we consider the non-zero mixing case, e.g., $\cos\theta =0.99$, we got the upper limit not only on $\sqrt{s}$ but also on $m_{\Phi}^{}$. 
Typically, we obtained the upper limit on $m_\Phi^{}$ in the non-zero mixing case to be ${\cal O}(1)$ TeV, 
but this can become stronger depending on the choice of the value of $\tan\beta$ and $M^2$. 
We also have considered the case with relaxed mass degeneracy, i.e., $m_A^{} \neq m_H^{}$. 
In particular for the case of $m_{H^\pm}=m_A=M$,  
we have observed that 
a somewhat larger mass region becomes available to the extra Higgs states $H$, $A$ or $H^\pm$ in the C2HDM with respect
to the E2HDM, the more so the smaller $f$. 
We have checked the similar behavior is seen in the case of $m_{H^\pm}=m_H^{}=M$. 
However, if we take $m_{H^\pm} = m_H$ and $M=m_A$, 
a larger value of $f$ gets a larger allowed parameter space. 
This is true irrespectively of whether we assume the additional doublet, with respect to the SM-like one, to be active or inert.

Hence, a thorough investigation of the Higgs mass patterns that may emerge at the LHC could enable us to find hints of a
C2HDM  hypothesis and to distinguish it from the E2HDM one. 
Also, from the analysis of the various scattering processes, one can infer the value of the compositeness scale $f$. 
This, however, requires the calculation of both production and decay rates of the various Higgs states, 
task which we postpone to a separate publication.

Finally, before closing, we would like to mention that our hybrid construction of the C2HDM, wherein
we are 
using the same form of the scalar potential as in the E2HDM except for the `kinetic' term that it is taken
to be the first order of a chiral expansion, makes it 
difficult to extract trustable hints about the
nature of the underlying dynamics of compositeness. In adopting such a choice for the scalar potential, we
are clearly inducing a model dependence in our approach. 
However, by choosing the most general CP-conserving 2HDM potential which is phenomenologically viable and 
 highlighting the parameter space regions where differences can be found between the E2HDM and C2HDM, our work
will  inform the choice of how to construct a realisation of a C2HDM (in terms of underlying gauge symmetries, their breaking patterns and the ensuing new bosonic and fermionic spectra) that is notably different from the E2HDM. In essence, 
our findings will  serve as a useful tool to take into account the constraints from perturbative unitarity in generic composite Higgs models 
with two Higgs-like doublets. Namely, if one calculates the CW potential in a given configuration of composite Higgs models, then 
all the parameters in the Higgs potential can be written in terms of those belonging to the composite sector 
(such as masses and couplings of strong resonances). 
Using such parameters, one can then easily apply the formulae of the unitarity bounds given in our paper to a C2HDM with a proper CW potential. 

\noindent
\section*{Acknowledgments}
\noindent 
All the authors are grateful to Shinya Kanemura for fruitful discussions. 
SDC would also like to thank Michele Redi, Andrea Tesi and Elena Vigiani for useful discussions.
KY is grateful to Kunio Kaneta for helpful discussions. 
The work of SM is financed in part through the NExT Institute and by the STFC Consolidated Grant ST/J000391/1.
This work was supported by a JSPS postdoctoral fellowships for research abroad (KY).
EY was supported by the Ministry of  National Education of Turkey.

\begin{appendix}

\section{Kinetic Term}

According to the prescription developed by Callan, Coleman, Wess, and Zumino~\cite{ccwz}, 
the kinetic Lagrangian in non-linear sigma models are expressed in Eq.~(\ref{ccwz1}). 
In this expression, each of $d_\mu$'s defined in Eq.~(\ref{ccwz2}) are calculated in the $SO(6)\to SO(4)\times SO(2)$ model by
\begin{align}
(d_\alpha^{\hat{1}})_\mu &= -\frac{\sqrt{2}}{f}\partial_\mu h_\alpha^{1}-\frac{g}{2f}[(h_\alpha^4-ih_\alpha^3)W_\mu^++(h_\alpha^4+ih_\alpha^3)W_\mu^-]\notag\\
&\quad\quad-\frac{\sqrt{2}g_Z^{}}{f}\left(\frac{1}{2}-\sin^2\theta_W\right)h_\alpha^2 Z_\mu 
-\frac{\sqrt{2}e^{}}{f}h_\alpha^2 A_\mu + {\cal O}(1/f^3), \\
(d_\alpha^{\hat{2}})_\mu &= -\frac{\sqrt{2}}{f}\partial_\mu h_\alpha^{2}-i\frac{g}{2f}[(h_\alpha^4-ih_\alpha^3)W_\mu^+-(h_\alpha^4+ih_\alpha^3)W_\mu^-]\notag\\
&\quad \quad +\frac{\sqrt{2}g_Z^{}}{f}\left(\frac{1}{2}-\sin^2\theta_W \right)h_\alpha^1 Z_\mu 
+\frac{\sqrt{2}e^{}}{f}h_\alpha^1 A_\mu+ {\cal O}(1/f^3), \\
(d_\alpha^{\hat{3}})_\mu &= -\frac{\sqrt{2}}{f}\partial_\mu h_\alpha^{3}+\frac{g}{2f}[(h_\alpha^2-ih_\alpha^1)W_\mu^++(h_\alpha^2+ih_\alpha^1)W_\mu^-]
-\frac{g_Z^{}}{\sqrt{2}f}h_\alpha^4 Z_\mu + {\cal O}(1/f^3), \\
(d_\alpha^{\hat{4}})_\mu &= -\frac{\sqrt{2}}{f}\partial_\mu h_\alpha^{4}+i\frac{g}{2f}[(h_\alpha^2-ih_\alpha^1)W_\mu^+-(h_\alpha^2+ih_\alpha^1)W_\mu^-]
+\frac{g_Z^{}}{\sqrt{2}f}h_\alpha^3 Z_\mu + {\cal O}(1/f^3).  
\end{align}
These expressions can be rewritten as 
\begin{align}
&i(d_\alpha^{\hat{1}})_\mu + (d_\alpha^{\hat{2}})_\mu
= -\frac{2}{f}\left[\partial_\mu \omega_\alpha^+ 
-i\frac{g}{\sqrt{2}}\phi_\alpha^0 W^+_\mu -ig_Z^{}\left(\frac{1}{2}-\sin^2\theta_W\right) \omega_\alpha^+ Z_\mu
-ie\omega_\alpha^+A_\mu\right]+ {\cal O}(1/f^3), \\
&-i(d_\alpha^{\hat{3}})_\mu + (d_\alpha^{\hat{4}})_\mu
=\frac{2}{f}\left[\partial_\mu \phi_\alpha^0 
-i\frac{g}{\sqrt{2}}\omega_\alpha^+ W^-_\mu +i\frac{g_Z^{}}{2}\phi_\alpha^0 Z_\mu \right]+ {\cal O}(1/f^3), 
\end{align}
where $\phi_\alpha^0= (h_\alpha^4-ih_\alpha^3)/\sqrt{2}$.

\end{appendix}

\end{document}